\documentclass[onecolumn,journal,12pt,draftclsnofoot]{IEEEtran}
\usepackage[T1]{fontenc} % optional T1 font encoding
\usepackage{cite}
\usepackage {graphicx}
\usepackage{amssymb}
\usepackage{amsmath}
\usepackage[ruled, linesnumbered]{algorithm2e}
\usepackage[caption=false, font=footnotesize]{subfig}
\usepackage{physics}
\usepackage[dvipsnames]{xcolor}
\usepackage{hyperref}
\usepackage{multirow}
\usepackage{units}

\usepackage{stfloats}

\usepackage{flushend}

\begin{document}
\title{Conditional Generative Adversarial Networks for Channel
	Estimation in RIS-Assisted ISAC Systems}

\author{Alice~Faisal,~\IEEEmembership{Graduate~Member,~IEEE,}
        Ibrahim~Al-Nahhal,~\IEEEmembership{Senior~Member,~IEEE,}
        Kyesan~Lee,~\IEEEmembership{Member,~IEEE,}
        ~Octavia~A.~Dobre,~\IEEEmembership{Fellow,~IEEE,}%
        ~and~Hyundong~Shin,~\IEEEmembership{Fellow,~IEEE}
%\thanks{Manuscript received 18 August 2024; revised 9 October 2024 and 9 January 2025; accepted 29 January 2025. This work was supported by the Canada Research Chairs Program under Grant CRC-2022-00187.}
\thanks{{
		Alice Faisal, Ibrahim Al-Nahhal, and Octavia A. Dobre (corresponding author) are with Memorial University, Canada. Octavia A. Dobre is also with Kyung Hee University, Korea. Kyesan Lee and Hyundong Shin (corresponding author) are with Kyung Hee University, Korea.
		%Alice Faisal, Ibrahim Al-Nahhal, and Octavia A. Dobre are with the Faculty of Engineering and Applied Science, Memorial University, Canada. Hyundong Shin is with Kyung Hee University, South Korea. 
 (e-mail: \href{mailto:afaisal@mun.ca}{afaisal@mun.ca}; \href{mailto:ioalnahhal@mun.ca}{ioalnahhal@mun.ca}; \href{mailto:odobre@mun.ca}{odobre@mun.ca};
 \href{mailto:kyesan@khu.ac.kr}{kyesan@khu.ac.kr};
  \href{mailto:hshin@khu.ac.kr}{hshin@khu.ac.kr})}}%
%\thanks{This work was supported by the Canada Research Chairs Program under Grant CRC-2022-00187.}
\thanks{Digital Object Identifier xx.xxxx/TCOMM.2025.xxxxxxx}%
}

% The paper headers
%\markboth{IEEE Transactions on Communications,~Vol.~xx, No.~x, January~2025}%
%{Shell \MakeLowercase{\textit{et al.}}: Bare Demo of IEEEtran.cls for IEEE Communications Society Journals}

\maketitle

\begin{abstract}
	Integrated sensing and communication (ISAC) technology has been explored as a potential advancement for future wireless networks, striving to effectively use spectral resources for both communication and sensing. The integration of reconfigurable intelligent surfaces (RIS) with ISAC further enhances this capability by optimizing the propagation environment, thereby improving both the sensing accuracy and communication quality. Within this domain, accurate channel estimation is crucial to ensure a reliable deployment. Traditional deep learning (DL) approaches, while effective, can impose performance limitations in modeling the complex dynamics of wireless channels. This paper proposes a novel application of conditional generative adversarial networks (CGANs) to solve the channel estimation problem of an {RIS-assisted} ISAC system. The CGAN framework adversarially trains two DL networks, enabling the generator network to not only learn the mapping relationship from observed data to real channel conditions but also to improve its output based on the discriminator network feedback, thus effectively optimizing the training process and estimation accuracy. {The} numerical simulations demonstrate that the proposed CGAN-based method improves the estimation performance effectively {compared} to conventional DL techniques. The results highlight {the} CGAN's potential to revolutionize channel estimation, paving the way for more accurate and reliable ISAC deployments.
\end{abstract}

%	The cGAN framework adversarially trains two DL networks to not only learn the mapping relationship from observed data to real channel conditions but also to adaptively refine the loss function, enhancing the training efficacy. 
	

\IEEEpeerreviewmaketitle

%__________________________________________________________________
\section{Introduction}

%The recent releases of the third generation partnership project (3GPP) mark significant steps forward in the evolution of the fifth-generation (5G) technology, referred to as 5G-Advanced. 

\IEEEPARstart{T}{he} {recent releases by the third generation partnership project (3GPP) represent crucial advancements in the evolution of fifth-generation (5G) networks, aiming at realizing the full potential  of 5G and bridging the transition to {the} sixth-generation (6G) networks. One of the key focus areas of 3GPP Release-19 is the integration of sensing capabilities into communication networks, which paves the way for new applications \cite{6G_1, 6G_2}. To this end, integrated sensing and communication (ISAC) is envisioned to play a key role in future generations of wireless networks by efficiently merging radar sensing with communication technologies within a single system. This convergence is expected to enhance spectrum utilization and reduce hardware costs compared to deploying separate systems with independent hardware for each function. ISAC is well-suited for a wide range of applications, such as drone operation, industrial automation, and health monitoring, where simultaneous data transmission and environmental sensing is needed. }
%	To this end, ISAC is  envisioned to enable innovative advancements in the upcoming generations of wireless networks by unlocking new possibilities in mobile communication and radar technologies \cite{ISAC_Magazine_2024}.}

%By leveraging ISAC, the {upcoming}  \textcolor{blue}{generations of wireless networks} will not only improve in terms of operational efficiency but also expand their capabilities to include advanced spatial awareness, allowing for innovations in both mobile communication and radar technologies . 

% ISAC allows for simultaneous data transmission and environmental sensing, making it ideal for diverse {applications, such} as drone operation, industrial automation, and smart cities, where  reliable data monitoring and communication are paramount. 

%	\IEEEPARstart{S}{ixth}-generation (6G) wireless networks are anticipated to revolutionize communication systems by delivering ultra-reliable low-latency communication, improved mobile broadband, and massive machine-type communications.  
%	These advanced features are essential for the next wave of transformative applications that demand high data rates and reliable communication across various services \cite{6G_1, 6G_2}. 

	 {ISAC has been investigated in the literature from various perspectives {to realize its full potential} \cite{ISAC_TR_2021, ISAC_TR2_2021, ISAC_TR3_2011, ISAC_secure_2021, ISAC_RA_2023, ISAC_RA2_2023, ISAC_Opt_2023, ISAC_Opt2_2023}. In particular, novel ISAC transceivers {were} {investigated through} {the development} of advanced beamforming and signal processing techniques \cite{ISAC_TR_2021, ISAC_TR2_2021, ISAC_TR3_2011}. Privacy and security concerns {were} addressed through the {design} of robust security protocols and privacy-preserving methods \cite{ISAC_secure_2021}.} Some works focused on resource allocation problems in ISAC-based systems to maintain effective sensing and communication (SAC) operations \cite{ISAC_RA_2023, ISAC_RA2_2023, ISAC_Opt_2023, ISAC_Opt2_2023}. Solving optimization problems in ISAC systems often involve complex trade-offs between enhancing signal quality and optimizing power consumption across both SAC tasks.  ISAC {faces additional challenges, such} as complex interference management issues that arise from simultaneously handling {SAC} tasks. Furthermore, the overlapping use of the spectrum for both functions can lead to signal contamination, necessitating advanced signal processing techniques to effectively distinguish between communication signals and sensing echoes. Moreover, the dual-use  can also introduce power management difficulties, as both {SAC} operations can be power-intensive. Addressing these challenges is crucial for realizing the potential of ISAC networks.

	Besides ISAC,  {the} reconfigurable intelligent surfaces (RIS) {have emerged as a potential technology to meet the demands of next-generation wireless networks.}
%	enhance wireless network capabilities and enable efficient wireless environments.
	 RIS consists of a two-dimensional array of low-cost passive elements that can be individually controlled to tune the phase, amplitude, and polarization of incoming signal. This enables  {RIS} to effectively mitigate interference, enhance signal strength at the  {receiver-side}, and extend the {wireless} communication coverage \cite{10316535}. Due to the unique properties  {of RIS and ISAC}, the integration of both technologies represents a significant advancement in the development of  {future} wireless networks. Notably, RIS can be deployed to help combat some of the ISAC system challenges. In particular, RIS can mitigate interference between {SAC} signals by enabling precise control over the propagation environment. Additionally, the passive nature of RIS elements can help reduce the overall energy consumption of {the} ISAC systems.
	 This capability enhances the {wireless} communication signal quality and minimizes the system power simultaneously. 
	
	%	 the inherent energy efficiency of {the} RIS can reduce the energy consumption of {the} ISAC systems due to their passive nature of manipulating electromagnetic waves without the need for power amplification. 
	
	 Recently, the joint integration of ISAC and RIS has been investigated in various {wireless} communication scenarios \cite{ISAC_secure2_2024, RIS_ISAC_2024, RIS_ISAC2_2024, RIS_ISAC3_2024, RIS_ISAC4_2023}. 
%	 These works focused on enhancing the overall communication system performance while considering the sensing metric constraints. 
	 For example, the {authors} in  \cite{ISAC_secure2_2024} and \cite{RIS_ISAC_2024} focused on optimizing {the base station (BS)} transmit precoding and RIS phase shifts to improve the radar estimation performance subject to the signal-to-interference-plus-noise ratio requirements of communication users. Furthermore, the work in \cite{RIS_ISAC2_2024} developed a beam training scheme that enables the BS to communicate with the users while also sensing targets.  {This approach distinguishes between the RIS and targets based on their mixed echoes, which is crucial to ensure seamless integration of RIS-assisted ISAC systems.} Since conventional methods often face limitations, either suffering from high computational complexity when using iterative approaches or poor performance when using heuristic approaches, some studies considered deep learning (DL) for resource allocation tasks \cite{ISAC_Opt3_2024, ISAC_Opt4_2024}.

	It is important to note that accurate channel state information (CSI) is needed for all the above designs. The aforementioned works generally assumed that CSI is available at the receiver side and only focused on the resource allocation and interference mitigation problems in RIS-assisted ISAC systems. However, in practical scenarios, obtaining accurate CSI poses a significant challenge due to the coexistence of {SAC} channels within the same system. This introduces the need of innovative channel estimation strategies that can effectively distinguish between the mixed signals in RIS-assisted ISAC systems, ensuring both accurate sensing data and reliable communication. Despite its importance, only a limited number of studies {addressed} the channel estimation challenges in RIS-assisted ISAC systems \cite{Yu_SU, Yu_Globecom, Yu_ELM, LIU202597}. In \cite{Yu_SU}, the authors focused on a single-RIS-assisted ISAC system and introduced a novel three-stage method to simplify the channel estimation process. The presented approach starts with direct channels estimation, followed by the estimation of reflected communication channel, and finally the reflected sensing channel. This framework utilizes two distinct convolutional neural network (CNN) models to accurately estimate the channels at the ISAC BS. Furthermore, the work in \cite{Yu_Globecom} considered a multi-user downlink RIS-assisted ISAC system and proposed two  deep neural network (DNN) models to estimate the SAC channels. Specifically, the first DNN is implemented at the ISAC BS for the sensing channel estimation, while the second is deployed at each downlink user for communication channel estimation. Finally, the work in \cite{Yu_ELM} considered a multi-user RIS-assisted ISAC system and proposed a two-stage framework to estimate the SAC channels, focusing sequentially on the direct and reflected links. The presented framework considered extreme learning machine to meet the system requirements with accelerated training speed.

	All the above works have benchmarked their performance against the least squares (LS) estimation technique. While commonly used, {the} LS estimation has several limitations including its sensitivity to noise, which can significantly degrade its performance in scenarios with low signal-to-noise ratio (SNR). Additionally, {the LS estimator} does not account for model non-linearities in the system, which can lead to biased and inefficient estimates in RIS-assisted ISAC environments. This gap points to the necessity for innovative approaches that  {enhance} the estimation performance efficiently. Therefore, this work  {applies the} conditional generative adversarial networks (CGANs) for channel estimation in RIS-assisted ISAC systems. Unlike regular DL methods that typically learn to map inputs to outputs in a supervised fashion, CGANs operate by training two models: a generator that creates data following the real data distribution, and a discriminator that learns to distinguish between the real and fake (i.e., generated) data. This adversarial process allows CGANs to generate accurate data and capture complex distributions, thereby enhancing the estimation and generalization capabilities over standard DL approaches. To the best of the authors' knowledge, this is the first work to consider employing CGANs for channel estimation tasks in {multi-user} RIS-assisted ISAC systems. The contributions {of this paper} are summarized as follows:
	\begin{itemize}
		\item A CGAN-based estimation framework, incorporating two distinct DNN architectures, is devised to estimate the SAC channels for  {a multi-user} RIS-assisted ISAC system. One DNN operates at the ISAC BS to estimate the sensing channel, while the other  {one} is deployed  {at} each downlink user to estimate the cascaded communication channel.
		\item A custom loss function is carefully designed for training {the proposed CGAN-based estimator} to ensure accurate and robust channel estimation performance even in environments with low SNRs. 
		\item The proposed CGAN framework operates as a minimax two-player game, where the generative and discriminative models continuously compete and evolve during the training process. This dynamic interaction leads to a generative model capable of producing channel samples that mimic real distribution patterns. This advanced approach enables the CGAN to exhibit exceptional adaptability and scalability, which is essential for RIS-assisted ISAC systems. 
		\item The proposed CGAN-based channel estimation approach demonstrates strong generalization capabilities, where it achieves robust performance at SNR ranges that are not considered during the training phase. This robustness indicate that the model effectively eliminates the need for  {the SNR} estimation stage in practical deployments.
		\item The computational complexity of the proposed CGAN approach is evaluated based on the number of real additions and multiplications. The numerical results indicate that the proposed  algorithm achieves a  {complexity comparable} to the DL-based benchmark estimator, which satisfies the low-cost deployment demands of RIS-assisted ISAC systems.
		\item Extensive simulations are conducted to validate the estimation performance of the proposed approach. Numerical findings prove that the proposed method significantly outperforms {the conventional DL-based approaches} {in the literature} under a  range of SNR conditions and system parameters.
	\end{itemize}

	The remainder of this paper\footnotemark is organized as follows: The system model and problem formulation {are introduced in  Section \ref{Sec:System_Model}}. The proposed {channel} estimation approach is {presented} in Section \ref{Sec:Proposed_Approach}, and the computational complexity is analyzed in Section \ref{Sec:Complexity_Analysis}. {Finally, simulation results are shown in Section \ref{Sec:Simulation_Results} and conclusions are drawn in Section \ref{Sec:Conclusion}.}

	\footnotetext{Boldface uppercase and lowercase letters denote matrices and vectors, respectively. \(\mathbb{C}\) and  \(\mathcal{CN}\) stand for a complex-valued variable and a complex-valued normally distributed random variable, respectively.  \(\Re \{\cdot\}\) and \(\Im \{\cdot\}\) denote the real and imaginary components of a variable, respectively. \(\mathbb{E} \{\cdot\}\) denotes the expectation operation. The operators \((\cdot)^{\mathrm{H}}, \mathrm{vec}[\cdot], (\cdot)^{-1}, \| \cdot \|_2\), and \(\| \cdot \|_F\) represent the Hermitian, vectorization, inverse, second order norm, and Frobenius norm of their arguments, respectively. \(\text{diag}\{\mathbf{x}\}\)  returns a matrix whose diagonal consists of the elements of \(\mathbf{x}\).}
		
%		 \(\mathcal{N}_{a}^{b} = \{a, a + 1, \ldots, b\}\) represents the index set from integer \(a\) to \(b\), and \(a < b\).}
		
%		(\cdot)^{\dagger}

%		 \(\sum \{\cdot\}\) returns the summation of all elements' values of a vector.}

%	Consequently, the generative network functions as a channel sample generation tool, the learned network parameters effectively represent the channel distribution characteristics.
	
	%	Moreover, as these studies where the first to consider the channel estimation problem in RIS-assited ISAC systems, they did not focus on reducing computational complexity relative to existing methods. This gap points to the necessity for innovative approaches that does not only enhance performance but are also computationally efficient. Therefore, 

	%	This integration marks a significant step towards more intelligent and adaptive wireless systems that can dynamically optimize their performance based on comprehensive environmental understanding.
	
%__________________________________________________________________
\section{System Model}~\label{Sec:System_Model}

	\begin{figure}[!t]
		\centering
		\subfloat[]{
			\includegraphics[width=0.40\textwidth]{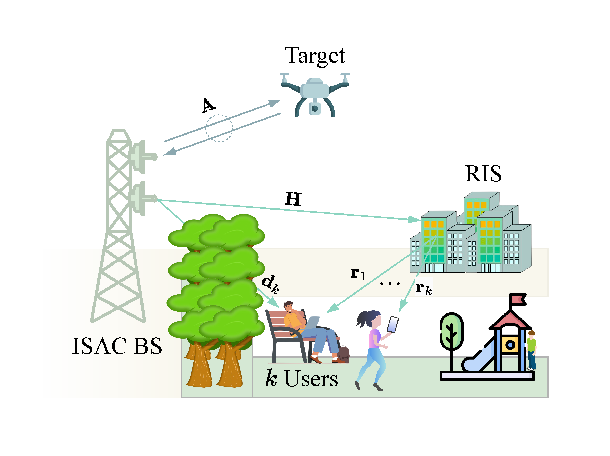}
			\label{Fig:System_a}
		}
		\hfil
		\subfloat[]{
			\includegraphics[width=0.48\textwidth]{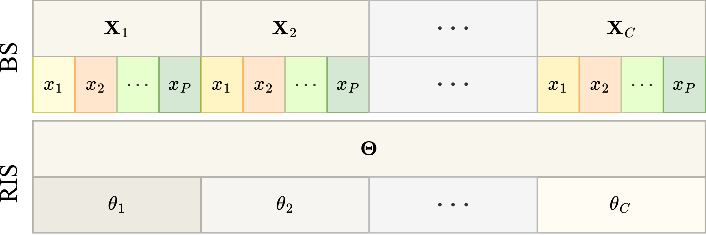}
			\label{Fig:System_b}
		}
		\caption{RIS-assisted ISAC system. (a) System model, (b) Pilot protocol.}
		\label{Fig:System}
	\end{figure}
	
	Consider an RIS-assisted {multiple-input single-output} (MISO) ISAC system as illustrated in Fig. \ref{Fig:System_a}, where the RIS facilitates the communication between the ISAC BS and $K$ downlink user {equipments (UEs)} and the BS communicates with a target for sensing purposes. The BS has $M$ transmit antennas
	and one receive antenna, and each UE is equipped with one receive antenna. The RIS consists of $N$ programmable reflecting elements. For target sensing, the BS sends radar signals towards the target and receives the echo signals through the BS-target-BS channel, denoted by $\mathbf{A} \in \mathbb{C}^{M\times M}$. The channel coefficients of {the} BS-RIS, RIS-UE$_{k}$, and BS-UE$_{k}$ are represented as $\mathbf{H} \in \mathbb{C}^{M\times N}$, $\mathbf{r}_k \in \mathbb{C}^{N\times 1}$, and $\mathbf{d}_k \in \mathbb{C}^{M\times 1}$, respectively. {Note that all the SAC channels are assumed to be flat-fading considering the narrowband transmission \cite{ch1,ch2,ch3,ch4}.} Given that the ISAC BS operates in full-duplex mode, transmitting and receiving signals simultaneously, it experiences self-interference (SI); {the SI channel is represented as $\mathbf{S} \in  \mathbb{C}^{M\times M}$.}

	A pilot transmission policy is developed to estimate the SAC channels in the {multi-user} RIS-assisted   ISAC system, as shown in Fig. \ref{Fig:System_b}. The ISAC BS transmits pilot sequences in $C$ sub-frames, {in which} each sub-frame is divided into $P$ time slots. The pilot signal of the ISAC BS at the $p$-th time slot is defined as $\mathbf{x}_p \in \mathbb{C}^{M\times 1}$. To this end, the pilot matrix in sub-frame $c$ is represented as $\mathbf{X}_c = \left[\mathbf{x}_1, \mathbf{x}_2, \cdots, \mathbf{x}_P \right] \in \mathbb{C}^{M\times P}$. It is {worth} noting that the ISAC BS transmits the same pilot sequences (i.e., denoted as $\mathbf{X}$) in each sub-frame. Consequently, the RIS {phase shifts} {remain unchanged} within a single sub-frame, {being denoted} by $\boldsymbol{\theta}_c \in \mathbb{C}^{N\times 1}$. The corresponding phase shift matrix is {represented by} $\boldsymbol{\Theta} = \left[\boldsymbol{\theta}_1, \boldsymbol{\theta}_2, \cdots, \boldsymbol{\theta}_C \right] \in \mathbb{C}^{N\times C}$. {Let $P = M$ and $C = N$ to accommodate the necessity for low pilot overhead. This design allows to effectively capture the necessary channel information without requiring additional pilot resources, which significantly reduces redundancy and prevents excessive pilot usage.} Furthermore, both $\mathbf{X}$ and $\boldsymbol{\Theta}$ are {modeled as a discrete} Fourier transform (DFT) matrix, expressed as

%	This design choice ensures that the number of pilot time slots and sub-frames matches the number of antennas and reflecting elements, respectively, which helps minimize the total pilot overhead. By aligning the pilot transmission duration with the dimensions of the system, we effectively capture the necessary channel information without requiring additional pilot resources. This approach significantly reduces redundancy and prevents excessive pilot usage, making it efficient for large-scale ISAC systems with a minimal impact on communication and sensing performance.}
%	
	\begin{equation}
		\mathbf{X} = 
		\begin{bmatrix}
			1 & 1 & \cdots & 1 \\
			1 & X^{1} & \cdots & X^{P - 1} \\
			\vdots & \vdots & \ddots & \vdots \\
			1 & X^{M - 1} & \cdots & X^{(M - 1)(P - 1)}
		\end{bmatrix},
	\end{equation}
	
	\noindent where $X^{(m, p)} = \frac{1}{\sqrt{M}} e^{{j \frac{2 \pi}{M} mp}}$ is the $(m,p)$-th entry of $\mathbf{X}$. Modeling $\mathbf{X}$ as a DFT matrix helps combat the interference between the SAC signals and distinguish between multi-user signals. Furthermore, it has been demonstrated that designing $\boldsymbol{\Theta}$ as a DFT matrix helps to boost the power of the received signal at UEs and ensure accurate channel estimation \cite{DFT_Proof}.
	
	To this end, the received signal at the $k$-th downlink UE for sub-frame $c$ and time slot $p$ is expressed as 
	
	\begin{equation}
		\label{Eq:RX_signal_UE_1}
		y^{\text{UE}}_{k,c,p} = \left(\mathbf{r}_k^H\text{diag}\{\boldsymbol{\theta}_c^H\} \mathbf{H}^H + \mathbf{d}^H_k \right) \mathbf{x}_p + n_{k,c,p}.
	\end{equation}
	
	\noindent Here, $n_{k,c,p}\sim\mathcal{CN}(0,\sigma^2)$ {is} the {complex} additive white Gaussian noise {(AWGN)} with zero-mean and variance $\sigma^2$. {The RIS phase shifts are expressed as \cite{RIS_Principles}}

	 \begin{equation}
	 	\begin{split}
		 	\boldsymbol{\theta}_c = 	\left[\beta_{c}e^{j\varphi_{c,1}},\beta_{c} e^{j\varphi_{c,2}},\cdots,\beta_{c} e^{j\varphi_{c, N}}\right]^T, \\ \varphi_{c, n} \in [0,2\pi), \hspace{2mm} \beta_{c} \in \left[0,1\right],
	 	\end{split}
	 \end{equation}
	 
	 \noindent where $\beta_{c} \in \left[0,1\right]$ represents the {RIS element} amplitude. The direct link is assumed to be blocked by obstacles and the reflected link is considered to support the communication. Here, $\text{diag}\left\{.\right\}$ transforms a vector into a diagonal matrix. This transformation implies the following
	
	\begin{equation}
		\text{diag}\left\{\boldsymbol{\theta}_c\right\} \mathbf{r}_k = \left [{\theta}_1 {r}_1, {\theta}_2 {r}_2 , \cdots, {\theta}_N {r}_N \right].
	\end{equation}
	
	\noindent Similarly, constructing $\text{diag}\{\boldsymbol{r}_k\} $ and multiplying it by $\boldsymbol{\theta}_c$ would produce the same result, {confirming the equality $\text{diag}\{\boldsymbol{\theta}_c\} \mathbf{r}_k = \text{diag}\{\mathbf{r}_k\} \boldsymbol{\theta}_c$ \cite{Horn_Johnson_1985}.} Using this property, the cascaded reflected channel of {the} BS-RIS-UE$_k$ is expressed as 
	
	\begin{equation}
		\mathbf{G}_k = \mathbf{H} \hspace{0.5mm} \text{diag}\{\mathbf{r}_k\} \in \mathbb{C}^{M\times N}.
	\end{equation}
	
	\noindent Therefore, \eqref{Eq:RX_signal_UE_1} can be formulated as 
	
	\begin{equation}
		\label{Eq:RX_signal_UE_2}
		y^{\text{UE}}_{k,c,p} = \boldsymbol{\theta}_c^H \mathbf{G}_k^H \mathbf{x}_p + n_{k,c,p}.
	\end{equation}

	On the other hand, the received signal at the ISAC BS is expressed as 
	
	\begin{equation}
		\label{Eq:RX_signal_BS}
		\mathbf{y}^{\text{BS}}_{c,p} = \underbrace{\mathbf{A}^H \mathbf{x}_p}_{\text{Sensing signal}} + \underbrace{\mathbf{S}^H \mathbf{x}_p}_{\text{SI}} + \mathbf{n}_{c,p},  
	\end{equation}
	
	\noindent where $\mathbf{n}_{c,p}\sim\mathcal{CN}(0,\sigma^2 \mathbf{I}_M)$ denotes the complex {AWGN} with zero-mean and variance $\sigma^2$. Here, $\mathbf{I}_M$ represents an identity matrix with size $M$. Since the propagation environment between the ISAC BS transmitting and receiving antennas is presumed to be stable \cite{SI_Proof}, the SI channel, $\mathbf{S}$, can be pre-determined at the ISAC BS. This allows for the compensation of any residual SI in \eqref{Eq:RX_signal_BS} prior to the {SAC} estimation process. 
%		\textcolor{blue}{Without loss of generality, the time slot duration (i.e., pilot symbol duration), denoted as $T_P$, is assumed to be significantly shorter than the channel coherence time, $T_{\text{coh}}$, but longer than the propagation delays of both the SAC and residual SI signals \cite{3gpp2016eutra, coh}. This assumption guarantees that the SAC channels remain constant throughout the estimation phase, while also enabling synchronization of the received SAC and residual SI signals.} 
%	\textcolor{blue}{Without loss of generality, the time slot duration (i.e., pilot symbol duration), denoted as $T_P$, is assumed to be significantly shorter than the channel coherence time, $T_{\text{coh}}$, but longer than the propagation delays of both the SAC and residual SI signals \cite{3gpp2016eutra, coh}. This assumption guarantees that the SAC channels remain constant throughout the estimation phase, while also enabling synchronization of the received SAC and residual SI signals.} 
	The estimation problems of the SAC channels in \eqref{Eq:RX_signal_UE_2} and \eqref{Eq:RX_signal_BS} are challenging. In what follows, we propose a novel approach to estimate both $\mathbf{G}_k$ and $\mathbf{A}$ based on the pilot protocol in Fig. \ref{Fig:System_b}.
	
%	\begin{equation}
%		\label{Eq:Transmitter_Signals}
%		\mathbf{x} =
%		\begin{cases}
%			\begin{bmatrix}
%				x & x & \cdots & x
%			\end{bmatrix}^T
%			& \text{Diversity}, \\[3mm]
%			\begin{bmatrix}
%				x_1 & x_2 & \cdots & x_{N_t}
%			\end{bmatrix}^T
%			& \text{Multiplexing}, \\[3mm]
%			\begin{bmatrix}
%				0 & \cdots & x_j & \cdots & 0
%			\end{bmatrix}^T
%			& \text{Spatial Modulation},
%		\end{cases}
%	\end{equation}

%__________________________________________________________________
\section{Proposed Estimation Approach}~\label{Sec:Proposed_Approach}	

	In traditional DL approaches for channel estimation, performance can saturate or generalize poorly when dealing with noisy or complex data environments, due to their dependence on direct mappings from input to output data. To address these limitations, {the} proposed method leverages GANs, which integrate a network feedback to refine the generation process. This adversarial setup not only enhances the ability to handle diverse and noisy datasets but also improves the network capability to generate realistic channel {estimates} that closely mimic actual channel conditions, ensuring robust performance across various scenarios. This section provides an overview {of GANs} and {then details} the proposed framework for the channel estimation problem in RIS-assisted ISAC systems. 
%------------------------------------------------------------------
	\subsection{Overview of Generative Adversarial Networks}
	
		GANs are a novel class of DL frameworks introduced in \cite{GAN_1}, consisting of two {DNNs:} the generator and discriminator. In this framework, the generator strives to produce data that cannot be distinguished from the original dataset. The discriminator, on the other hand, tries to classify the given data as real or fake (i.e., generated samples). Both networks engage in a continuous adversarial training process, where the discriminator learns to identify generated data more accurately and the generator learns to better generate data samples given the mutual feedback. This process continues until the generator learns to produce real-like samples that the discriminator fails to identify as fake. 
		
		GANs and their variants have been utilized {across a range} of applications {beyond the realm of wireless communications}. It has been considered to generate synthetic human faces  and transforming real-world scenery images into styles of famous paintings \cite{GAN_image}. Additional examples include enhancing image resolution, and generating music, audio, and video \cite{GAN_music, GAN_audio, GAN_resolution}. In the field of wireless communications, GANs have emerged as powerful tool for various applications \cite{GAN_magazine}. In particular, GANs has been used to generate synthetic data that {augment} existing datasets in situations where collecting real wireless communication data is challenging. This is particularly {useful for} tasks such as signal classification, mitigating wireless jamming attacks, and spectrum sensing \cite{GAN_secure, GAN_calssification, GAN_Sensing}. {GANs have} been further considered for optimizing wireless systems, {where they learn} to generate accurate resource allocation decisions by modeling the network environment and the continuous feedback from the discriminator \cite{GAN_optimization}. Overall, GANs contribute significantly to advancements in network design, optimization, and data generation in wireless communications, driving forward innovations that enhance the efficiency and reliability of wireless systems.
		
		In traditional GANs, the generator, $G$, takes as an input random noise vector, $\mathbf{z}$, typically drawn from a Gaussian or uniform distribution, $d(\mathbf{z})$. It outputs data that {resemble} the target data on which the model was trained, denoted by $G(\mathbf{z})$. The discriminator takes the target (i.e., real data) as an input, $\mathbf{x}$, with distribution  $d(\mathbf{x})$, and the generated data,  $G(\mathbf{z})$. The output of the discriminator is a binary classification, {$D(\mathbf{x})$ and  $D(G(\mathbf{z})) \in \{0, 1\}$}, where the goal is to determine whether the given {data are real or} fake. The objective function of GANs is represented as 
		
		\begin{equation}
			\label{Eq: GAN_obj}
			\min_G \max_D \mathbb{E}_{\mathbf{x}\sim d(\mathbf{x})}\left[\log(D(\mathbf{x}))\right] + \mathbb{E}_{\mathbf{z}\sim d(\mathbf{z})}\left[\log(1- D(G(\mathbf{z})))\right].
		\end{equation}
		
		\noindent The first term in \eqref{Eq: GAN_obj} represents the expected value of the discriminator output (i.e., probability that $\mathbf{x}$ is real) over all real data samples, $\mathbf{x}$. The discriminator focuses on maximize this term by assigning higher probabilities to real data, thus maximizing $\log(D(\mathbf{x}))$. Furthermore, the discriminator aims to maximize the value of the second term by minimizing $D(G(\mathbf{z}))$ (i.e., probability that input is generated). On the other hand, the generator aims to maximize $D(G(\mathbf{z}))$ by convincing the discriminator that the generated input is real. 
		
		Both the generator and discriminator are updated through an iterative process, each using a distinct loss function derived from the overall  objective function in \eqref{Eq: GAN_obj}. This training process involves alternating updates to the discriminator and the generator with the goal of optimizing their respective loss functions. The loss functions of the discriminator and generator are {respectively} given  as 
		
		\begin{equation}
		L_D = -\frac{1}{2} \sum_{j=1}^{b} \left[ \log D(\mathbf{x}_j) + \log (1 - D(G(\mathbf{z}_j))) \right],
		\end{equation}
		\noindent and 
		\begin{equation}
			L_G = -\frac{1}{b} \sum_{j=1}^{b} \log D(G(\mathbf{z}_j)),
		\end{equation}
		
		\noindent where b is the batch size.

%			In particular, the generator learns to create data that can fool the discriminator but does not necessarily keep needed features or characteristics of the target. 
%		This can lead to producing a limited diversity of outputs and lack of ability to interpret how changes in the input noise vector affect the outputs, which is a crucial feature in complicated environments, such as wireless communication systems.

		Given the above concept, GANs have made significant progress in the field of generative modeling, offering the ability to produce highly realistic samples across various domains. However, despite their successes, {traditional GANs face a primary challenge} that can affect their applicability to wireless systems. The main issue is the lack of control over the generated data. {In particular, the generator learns to create data that can fool the discriminator but does not necessarily retain the essential features or characteristics of the target. This can result in a lack of diversity in the outputs, making it difficult to capture the complexity of channel variations and the specific structure of wireless communication signals. Furthermore, traditional GANs have limited interpretability, where changes in the input noise vector do not directly correspond to meaningful variations in the generated data, which is a critical requirement in complex environments, such as wireless communication systems.}

%		In CGANs, the generator and discriminator models are conditioned on the class label or target data, allowing the generator to generate samples of a given sample distribution. This conditioning allows the model to generate data more effectively, leading to improved diversity, stability, and faster training. 

		{To address these limitations, CGANs have been specifically developed to incorporate additional information that guides the data generation process \cite{mirza2014conditional}. In CGANs, the generator and discriminator models are conditioned on auxiliary information, such as class labels or specific data attributes. This conditioning allows the generator to produce samples that align with the provided conditions, ensuring that the generated outputs maintain the necessary characteristics of the target data distribution. As a result, CGANs offer improved diversity, stability, and faster convergence in training.} {In the context of channel estimation, CGANs enable the generator to be conditioned on different channel states or environmental parameters, which is essential for accurately modeling the characteristics of the channel. This conditioning mechanism allows CGANs to adapt to variations in channel properties, leading to a more robust channel estimation process that can handle rapidly changing communication environments. Unlike basic GANs, which rely solely on input noise, the proposed CGAN framework in this work uses received observations as conditional inputs, allowing it to learn and reproduce complex channel structures with higher accuracy}.  In what follows, we detail the proposed CGAN {channel} estimation approach for RIS-assisted ISAC systems.

%		In the context of channel estimation, CGANs can be used by conditioning the network on different channel states or environmental conditions, which enables the network to capture channel characteristics rather than depending on a noisy input. This further leads to designing more robust communication systems that can adapt to rapidly changing conditions. In what follows, we detail the proposed CGAN {channel} estimation approach for RIS-assisted ISAC systems.

%------------------------------------------------------------------
	\subsection{Proposed CGAN-Based DL Framework}
		This section introduces a novel channel estimation approach utilizing CGANs. Initially, the configuration of the input-output pairs for the framework is carefully designed. Building on this, a CGAN-based {channel} estimation approach is developed to enhance {the} estimation performance.
		
		\subsubsection{Input-Output Pair Design}
			\paragraph{SAC Design}
				Consider the received signal in \eqref{Eq:RX_signal_UE_2} to construct the input-output pairs for estimating the communication channel. {In each sub-frame, the received signal at UE$_k$ across $P$ time slots is represented by}
				
				\begin{equation}
					\label{Eq:RX_signal_UE_F}
					\mathbf{y}^{\text{UE}}_{k,c} = \boldsymbol{\theta}_c^H \mathbf{G}_k^H \mathbf{X} + \mathbf{n}_{k,c},
				\end{equation}
				
				\noindent where $\mathbf{y}^{\text{UE}}_{k,c} = \left[y^{\text{UE}}_{k,c,1}, y^{\text{UE}}_{k,c,2}, \cdots, y^{\text{UE}}_{k,c,P} \right] \in \mathbb{C}^{1\times P}$ and $\mathbf{n}_{k,c} = \left[n_{k,c,1}, n_{k,c,2}, \cdots, n_{k,c,P} \right] \in \mathbb{C}^{1\times P}$. Based on this, the communication data design is given as 
				
				\begin{equation}
					\begin{split}
						\mathbf{R}^{\text{UE}_{k}} =& \left[\Re{	\mathbf{y}^{\text{UE}}_{k,1}, 		\mathbf{y}^{\text{UE}}_{k,2}, \cdots, 	\mathbf{y}^{\text{UE}}_{k,C}}, \right. \\ & \left. \Im{\mathbf{y}^{\text{UE}}_{k,1}, \mathbf{y}^{\text{UE}}_{k,2}, \cdots, 	\mathbf{y}^{\text{UE}}_{k,C}}\right]^T.
					\end{split}
				\end{equation}
				
				\noindent The corresponding output (i.e., target) is the ground truth channel, $\mathbf{G}_k$, given as
				
				\begin{equation}
					\mathbf{O}^{\text{UE}_{k}} = \left[\Re{\text{vec}[{\mathbf{G}_k}]}, \Im{\text{vec}[{\mathbf{G}_k}]}\right]^T.
				\end{equation}
				
%				\noindent where $\text{vec}\{.\}$ denotes the operation of converting a matrix into a vector.
				
				On the other hand, to construct the input-output pairs for estimating the sensing channel, consider the received signal in \eqref{Eq:RX_signal_BS}. {Similarly, the received signals at the ISAC BS are stacked across each sub-frame, $C$, leading to} 
				
				\begin{figure*}[!t]
					\centering
					\subfloat[]{
						\includegraphics[width=0.9\textwidth]{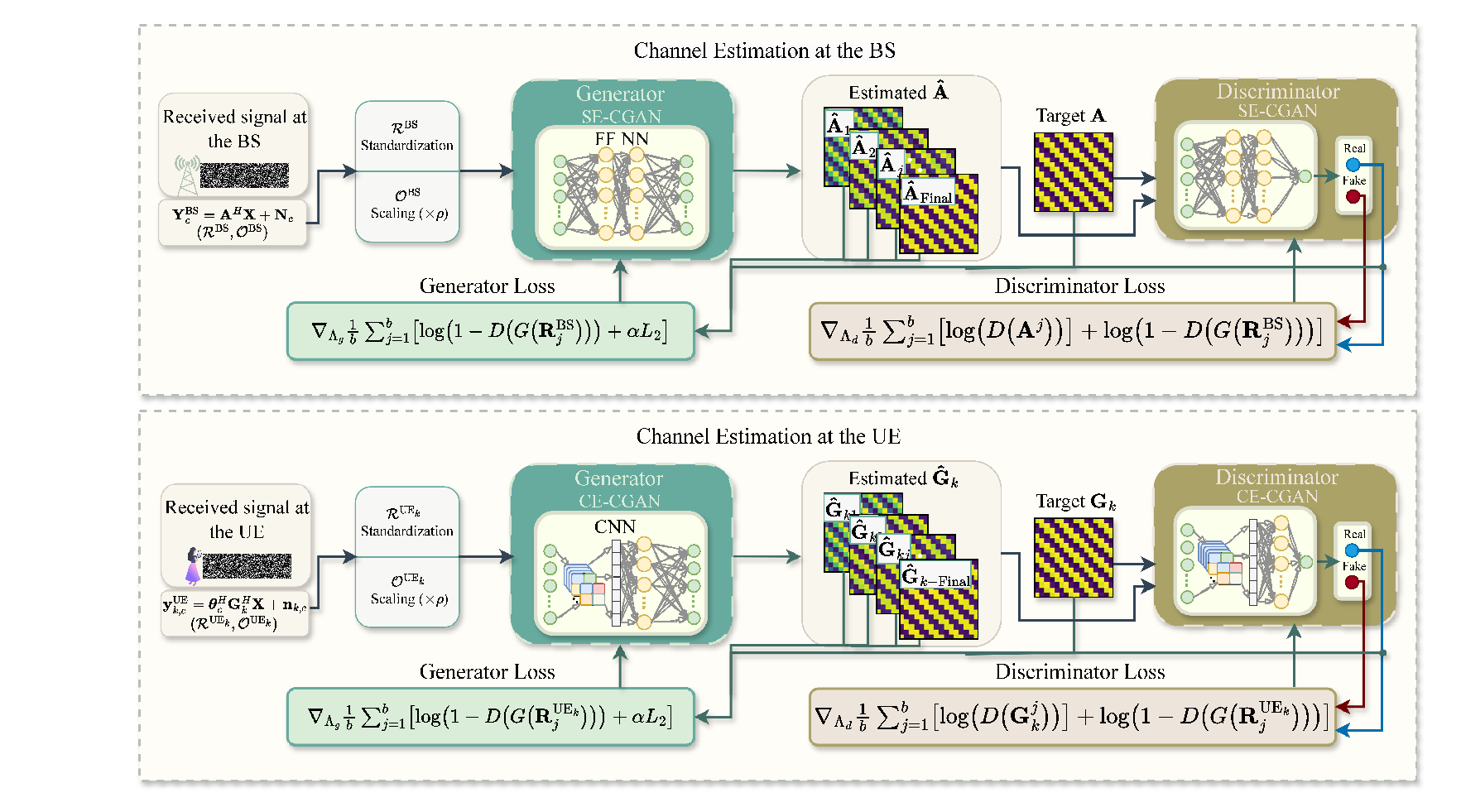}
						%				\caption{Training phase.}
						\label{Fig:CGAN_proposed_a}
					}
					\hfil
					\subfloat[]{
						\includegraphics[width=1\textwidth]{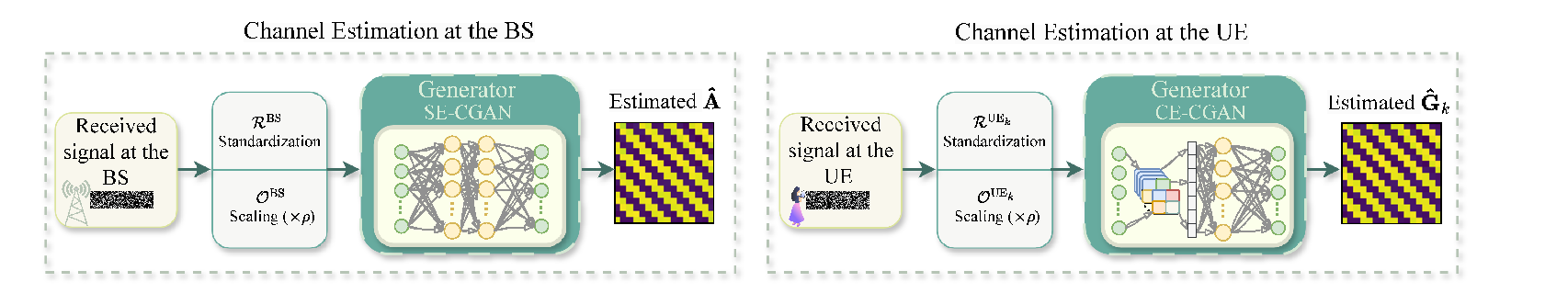}
						%				\caption{Testing phase.}
						\label{Fig:CGAN_proposed_b}
					}
					\caption{{Proposed CGAN-based channel estimation framework for RIS-assisted ISAC system. (a) Offline training. (b) Online testing.}}
					\label{Fig:CGAN_proposed}
				\end{figure*}

				\begin{equation}
					\label{Eq:RX_signal_BS_F}
					\mathbf{Y}^{\text{BS}}_{c} = \mathbf{A}^H \mathbf{X} + \mathbf{N}_{c}, 
				\end{equation}
				\noindent where $\mathbf{N}_{c} = \left[\mathbf{n}_{c,1}, \mathbf{n}_{c,2}, \cdots, \mathbf{n}_{c,P} \right] \in \mathbb{C}^{M\times P}$ and $\mathbf{Y}^{\text{BS}}_{c} = \left[\mathbf{y}^{\text{BS}}_{c,1}, \mathbf{y}^{\text{BS}}_{c,2}, \cdots, \mathbf{y}^{\text{BS}}_{c,P} \right] \in \mathbb{C}^{M\times P}$. To this end, the sensing data is generated as 
				\begin{equation}
					\begin{split}
						\mathbf{R}^{\text{BS}} =& \left[ 
						\Re{\text{vec}\left[\mathbf{Y}^{\text{BS}}_{1}, 	\mathbf{Y}^{\text{BS}}_{2}, \cdots, 	\mathbf{Y}^{\text{BS}}_{C}\right]}, \right. \\ & \left.
						\Im{\text{vec}\left[\mathbf{Y}^{\text{BS}}_{1}, 	\mathbf{Y}^{\text{BS}}_{2}, \cdots, 	\mathbf{Y}^{\text{BS}}_{C}\right]} \right]^T.
					\end{split}
				\end{equation}
				
				\noindent Accordingly, the output is the ground truth channel, $\mathbf{A}$, given as
				
				\begin{equation}
					\mathbf{O}^{\text{BS}} = \left[\Re{\text{vec}[{\mathbf{A}}]}, \Im{\text{vec}[{\mathbf{A}}]}\right]^T.
				\end{equation}
				
				\noindent {It is worth noting that ground truth channels are used solely for the training phase. During training, the availability of the true channel values enables the CGAN network to learn the underlying channel generation process effectively. However, in the testing phase, the only available information is the observation, represented by the received signal at the BS or UE. In this phase, the trained CGAN relies on this received signal to generate an estimate of the channels. This ensures that the CGAN model generalizes well to real-world scenarios, where the true channels are unknown.} Note that the generated {dataset} will not be used for direct input-output relationship as the case with regular NNs. The following sub-section will detail the working principle of the proposed CGAN {channel} estimation framework. 
				
			\paragraph{{Dataset} Generation}
			\label{subsec: data_gen}
				The training samples are produced by utilizing $Q$ received signals (i.e., each of the  communication and sensing signals in \eqref{Eq:RX_signal_UE_F} and \eqref{Eq:RX_signal_BS_F}, {respectively}), along with $V - 1$ duplicates of the $q$-th signal. The duplicates  are created by adding synthetic {AWGN} to the original channel according to $\text{SNR}_{\mathbf{A}/\mathbf{G}_{k}} = \frac{p_{{\mathbf{A}/\mathbf{G}_{k}}}}{\sigma^2_{\mathbf{A}/\mathbf{G}_{k}}}$, where $p_{{\mathbf{A}/\mathbf{G}_{k}}}$ and $\sigma^2_{\mathbf{A}/\mathbf{G}_{k}}$ denote the channel power and synthetic noise, respectively \cite{Yu_Globecom}. This dataset includes both noise-free original signals and their noisy versions to enhance the robustness and performance of the CGAN estimation model. Accordingly, the dataset for communication and sensing can be respectively represented as 
				
				\begin{equation}
					\begin{split}
						(\mathcal{R}^{\text{UE}_{k}},\mathcal{O}^{\text{UE}_{k}}) = 	\left\{ \left( \mathbf{R}^{\text{UE}_{k}}_{(1,1)}, \mathbf{O}^{\text{UE}_{k}}_{(1)} \right), \left( \mathbf{R}^{\text{UE}_{k}}_{(1,2)}, \mathbf{O}^{\text{UE}_{k}}_{(1)} \right), \cdots, \right. \\ \left.
						\left( \mathbf{R}^{\text{UE}_{k}}_{(1,V)}, \mathbf{O}^{\text{UE}_{k}}_{(1)} \right), \left( \mathbf{R}^{\text{UE}_{k}}_{(2,1)}, \mathbf{O}^{\text{UE}_{k}}_{(2)} \right),  \cdots, \left( \mathbf{R}^{\text{UE}_{k}}_{(Q,V)}, \mathbf{O}^{\text{UE}_{k}}_{(Q)} \right)  \right\},
					\end{split}
				\end{equation}
				\noindent and 
				\begin{equation}
					\begin{split}
					(\mathcal{R}^{\text{BS}},\mathcal{O}^{\text{BS}}) = 	\left\{ \left( \mathbf{R}^{\text{BS}}_{(1,1)}, \mathbf{O}^{\text{BS}}_{(1)} \right), \left( \mathbf{R}^{\text{BS}}_{(1,2)}, \mathbf{O}^{\text{BS}}_{(1)} \right), \cdots, \right. \\ \left.
					\left( \mathbf{R}^{\text{BS}}_{(1,V)}, \mathbf{O}^{\text{BS}}_{(1)} \right), \left( \mathbf{R}^{\text{BS}}_{(2,1)}, \mathbf{O}^{\text{BS}}_{(2)} \right),  \cdots, \left( \mathbf{R}^{\text{BS}}_{(Q,V)}, \mathbf{O}^{\text{BS}}_{(Q)} \right)  \right\}.
					\end{split}
				\end{equation}
				
%		\begin{figure}[!t]
%			\centering
%			\includegraphics[width=0.50\textwidth]{Images/Proposed_Approach_v3.pdf}
%			\caption{Proposed CGAN-based estimation approach.}
%			\label{Fig:CGAN_proposed}
%		\end{figure}

		\subsubsection{Working Principle}
		The proposed CGAN-based estimation framework to estimate $\mathbf{A}$ and $\mathbf{G}_k$ is shown in Fig. \ref{Fig:CGAN_proposed} and detailed as follows 
		
			\paragraph{Offline Training}

				The generator {NN} aims to estimate the SAC channels based on the conditional input, {which includes} the received signal observations, $\mathbf{R}^{\text{BS}}$ or $\mathbf{R}^{\text{UE}_{k}}$. {Prior to being passed into the generator, these received signals undergo preprocessing steps to ensure that the input data has a consistent range and to improve the model stability. A standardization is performed on the \((q, v)\)-th input data, which can be mathematically {expressed as}}

				\begin{equation}
					\label{eq:std}
					{\begin{split}
						\mathbf{R}_{\text{std}}^{\text{BS}/\text{UE}_k}(q, v) = 	\frac{\mathbf{R}^{\text{BS}/\text{UE}_k}(q, v) - \mathbb{E}\{\mathbf{R}^{\text{BS}/\text{UE}_k}(q, v)\}}{\mathbb{D}\{\mathbf{R}^{\text{BS}/\text{UE}_k}(q, v)\}}, 
					\end{split}}
				\end{equation}
				
				\noindent {where $\mathbb{E}\{\mathbf{R}^{\text{BS}/\text{UE}_k}(q, v)\}$ and ${\mathbb{D}\{\mathbf{R}^{\text{BS}/\text{UE}_k}(q, v)\}}$ denote the mean and standard deviation of the input, respectively. Additionally, the estimation target is scaled to facilitate the learning process as follows}
					
					\begin{equation}
					{\label{eq:scale}
						\mathbf{O}_s^{\text{BS}/\text{UE}_k}(q,v) = \rho \mathbf{O}^{\text{BS}/\text{UE}_k}(q,v),}
					\end{equation}

					\noindent {where $\rho$ is the scaling factor.}

				{The discriminator {NN}, on the other hand, aims to  identify whether a {given input is real (i.e., labeled $1$) or fake (i.e., labeled $0$), as derived} from the generator output; this process is summarized as follows}

			\begin{equation}
			{	D(\mathbf{B}) = 
				\begin{cases}
					1, & \text{if } \mathbf{B} \in \{\mathbf{A}, 	\mathbf{G}_k\} \\ 
					0, & \text{if } \mathbf{B} \in 	\{\hat{\mathbf{A}}, \hat{\mathbf{G}}_k\} 
				\end{cases}}
			\end{equation}
				
				\noindent  {where $\mathbf{B}$ represents the input to the discriminator, either as ground truth channel matrices, $\mathbf{A}$ or $\mathbf{G}_k$, or generated estimates $\hat{\mathbf{A}}$ or $\hat{\mathbf{G}}_k$.}
				
				In this work, the overall objective function of GANs presented in \eqref{Eq: GAN_obj} is further improved by adding an $L_2$ term. {This modification guides the generator towards more {accurate estimates}, ensuring that the generated channel not {only deceives} the discriminator but also {closely approximates} the true channel values. Furthermore, the $L_2$ loss term serves as a regularizer, penalizing large deviations from the target values. This helps prevent overfitting and fosters a model that generalizes better to new, unseen data while maintaining robustness against noise in received signals. To this end, {\eqref{Eq: GAN_obj} becomes}}
				
				\begin{equation}
					\begin{split}
					\min_G \max_D \hspace{2mm} & \mathbb{E}\bigl[\log\bigl(D\bigl(\mathbf{A}, \mathbf{G}_{k}\bigr)\bigr)\bigr] + \\ & \mathbb{E}\bigl[\log\bigl(1- D\bigl(G\bigl(\mathbf{R}^{\text{BS}}\bigr), G\bigl(\mathbf{R}^{\text{UE}_{k}}\bigr)\bigr)\bigr)\bigr] + \alpha L_2,
					\end{split}
				\end{equation}
				
				\noindent where

				\begin{equation}
					\label{Eq: CGAN_Obj_F}
					{L_2 = \mathbb{E}\bigl[\bigl\| \mathbf{B} - G(\mathbf{R}) \bigr\|^2 \bigr].}
				\end{equation}
				
				\noindent  {Here, $\mathbf{B}$ {denotes} the SAC channels (i.e., $\mathbf{B} = \{\mathbf{A}, \mathbf{G}_{k}\}$), and $\mathbf{R}$ is the input data ($\text{i.e., } \mathbf{R} = \{\mathbf{R}^{\text{BS}}, \mathbf{R}^{\text{UE}_{k}}\}$). Furthermore, $\alpha$  a {weighting} factor that enhances the stability of the channel estimates by penalizing large deviations from the target outputs. The generator and discriminator are trained together in an adversarial manner, where each network contributes to the loss of the other. Specifically, the generator aims to create estimates that can fool the discriminator, while the discriminator aims to correctly distinguish between real and generated data. This adversarial training is repeated until convergence, where the generator learns to produce realistic estimates. Fig. \ref{Fig:CGAN_proposed_a} illustrates the proposed {CGAN} DNN architectures for the SAC channels estimation and summarizes the offline training process.}

				In particular, two different architectures are developed within the CGAN framework: one for sensing channel estimation (SE-CGAN) and the other {one} for communication channel estimation (CE-CGAN). The SE-CGAN network is designed as feedforward (FF) fully connected NN, while the CE-CGAN {as CNN} to handle the complicated input-output relationship associated with {the} cascaded communication channel. In particular, {the proposed} SE-CGAN starts with an input layer, an FF layer {(FFL)} and a batch normalization layer to stabilize the NN output. The output is then fed to the second FFL followed by the second batch normalization and an output layer. On the other hand, the CE-DNN consist of one convolutional layer {(CL)}, batch normalization layer, followed by a flatten layer, an FFL, and a second batch normalization layer layer at the end. The two hidden layers for each of {the} SE-CGAN and CE-CGAN networks use the \textit{LeakyReLU} activation function. {This is defined as follows}
				
 				\begin{equation}
					\text{LeakyReLU}(x) = 
					\begin{cases} 
						x & \text{if } x > 0 \\
						\gamma x & \text{otherwise}.
					\end{cases}
				\end{equation}
				
				\noindent {Here, \(\gamma\) is a small constant value, which is set to $0.2$.} Both CGANs utilize a discriminator architecture that is identical to their respective generators. The only difference {consists} in the output layer size; the generator outputs the estimated channels, whereas the discriminator outputs a single value {(i.e., $0$ or $1$)} to evaluate the input data. {The \textit{Adam} optimizer is adopted to update the generator and discriminator parameters  with a learning rate
				of $2 \times 10^{-4}$ and $2 \times 10^{-5}$, respectively \cite{kingma2014adam}}. {These learning rates were carefully chosen based on extensive hyperparameter tuning. During this process, various learning rates were tested to find an optimal balance between convergence speed and estimation accuracy. Lower learning rates helped achieve more stable performance. Selecting a slightly higher rate for the generator enabled faster adaptation to the changes in the discriminator feedback.} The minibatch transitions is set to size $16$ and $\alpha = 100$. The detailed architectures of the SE-CGAN and CE-CGAN are summarized in Table \ref{tab:network_architecture}.

			\paragraph{Online Testing}

				{The online testing phases for SAC channel {estimation are} shown in Fig. \ref{Fig:CGAN_proposed_b}. During the testing phase, only the generator network is used, with no access to the discriminator network. The generator, having been trained to generate accurate channel estimates, is now applied to the testing data to predict the channel matrices. The testing dataset, $\dot{\mathcal{R}}^{\text{BS}}$ and $\dot{\mathcal{R}}^{\text{UE}_{k}}$, first undergoes preprocessing to standardize the samples according to \eqref{eq:std}, denoted by $\bar{\mathbf{R}}^{\text{BS}}$ and $\bar{\mathbf{R}}^{\text{UE}_{k}}$, {and then is fed} to the generator network.} The estimated output is represented as
				\begin{equation}
					\begin{split}
						\hat{\mathbf{A}} &= \rho^{-1} G(\bar{\mathbf{R}}^{\text{BS}} ; \hat{\Lambda}) \\ \hat{\mathbf{G}}_{k} &= \rho^{-1} G(\bar{\mathbf{R}}^{\text{UE}_{k}}; \hat{\Lambda}),
					\end{split} 
				\end{equation}
				\noindent where $\hat{\Lambda}$ {represents} the hyperparameters of the trained CGAN network (i.e., SE-CGAN or CE-CGAN). The output of the generator is scaled by a factor of $\rho^{-1}$ to obtain the channel matrices, $\hat{\mathbf{A}}$ and $\hat{\mathbf{G}}_{k}$. The proposed CGAN-based channel estimation algorithm is detailed in Algorithm \ref{alg:cgan}, where $\Lambda_g$ and $\Lambda_d$ denote the generator and discriminator parameters, respectively.

%				Output the trained SE-CGAN, , at the ISAC BS
%				and the trained CE-CGAN, , at the downlink UE_$k$.
%				$G(\bar{\mathbf{R}}^{\text{BS}} ; \hat{\Lambda}_{g}) \\ \hat{\mathbf{G}}_{k}$  
%				
%				
%				$G(\bar{\mathbf{R}}^{\text{UE}_{k}}; \hat{\Lambda_{g}})$
%				
%				Output the trained SE-DNN, fS(· ; ˆΘS), at the ISAC BS
%				and the trained CE-DNN, fUk (· ; ˆΘUk ), at the downlink Uk.
		
				%\begin{algorithm}[!htb]
%	\caption{CGAN-based Channel Estimation Algorithm.}
%	\label{Alg:CGAN}
%	Initialize $\lambda \in (1, 4/3)$, $m \gets 1$, and $L_{\,\text{tot}} \gets 0$\;
%	Randomly choose $L \in \{0, 1, \dots, \left\lfloor m \right\rfloor\}$\;
%	Apply Grover's algorithm with $L$ iterations of $G$ which outputs $\alpha_o$\;
%	Update $L_{\,\text{tot}} \gets L_{\,\text{tot}} + L$\;
%	\eIf{$\alpha_o=\alpha_d \; \mathrm{or} \; L_{\mathrm{tot}} \ge \Big\lceil 4.5\sqrt{N} \Big\rceil$}{
%		Return $\alpha_o$\;
%	}{
%		Update $m \gets \min\left(\lambda m, \sqrt{N}\right)$\;
%		Go back to Step 2\;
%	}
%\end{algorithm}

\begin{algorithm}
	\caption{Training a {CGAN}}
	\label{alg:cgan}
		\textbf{Initialize} the generator $G$ and discriminator $D$ with random weights, {batch size $b$}, learning rates, and training epochs $\epsilon$;\
		
		\textbf{Generate}  $(\mathcal{R}^{\text{UE}_{k}},\mathcal{O}^{\text{UE}_{k}})$ and $(\mathcal{R}^{\text{BS}},\mathcal{O}^{\text{BS}})$ according to Sec. \ref{subsec: data_gen};\
		
		\textbf{Pre-process} $(\mathcal{R}^{\text{UE}_{k}},\mathcal{O}^{\text{UE}_{k}})$ and $(\mathcal{R}^{\text{BS}},\mathcal{O}^{\text{BS}})$;\
		
		\For{epoch $i = 1$ to $\epsilon$}{
			\For{batch $j=1$ to $B$}
			{\textbf{Sample} batch of samples from $\mathbf{R}^{\text{UE}_{k}}$ or $\mathbf{R}^{\text{BS}}$\;
			\textbf{Sample} batch of samples from $\mathbf{O}^{\text{UE}_{k}}$ or $\mathbf{O}^{\text{BS}}$\;
			\textbf{Generate} samples from the generator $G(\mathbf{R}^{\text{UE}_{k}})$ or $G(\mathbf{R}^{\text{UE}_{k}})$\;
			\textbf{Evaluate} generated samples $D(G(\mathbf{R}^{\text{UE}_{k}}))$ or $D(G(\mathbf{R}^{\text{UE}_{k}}))$\;
			\textbf{Evaluate} target samples $D(\mathbf{O}^{\text{UE}_{k}})$ or $D(\mathbf{O}^{\text{BS}})$\;
			\textbf{Update} $D$ by ascending its stochastic gradient:
			\begin{equation}
			\begin{split}
				\nabla_{\Lambda_d} & \frac{1}{b} \sum_{j=1}^b 	\bigl[\log\bigl(D\bigl(\mathbf{A}^{j}, \mathbf{G}^{j}_{k}\bigr)\bigr)\bigr] + \bigr. \\ \bigl.  & \log\bigl(1- D\bigl(G\bigl(\mathbf{R}_{j}^{\text{BS}}\bigr), G\bigl(\mathbf{R}_{j}^{\text{UE}_{k}}\bigr)\bigr)\bigr)\bigr]
			\end{split}
			\end{equation}

			\textbf{Update} $G$ by descending its stochastic gradient:
			\begin{equation}
				\nabla_{\Lambda_g} \frac{1}{b} \sum_{j=1}^b \bigl[ \log\bigl(1- D\bigl(G\bigl(\mathbf{R}_j^{\text{BS}}\bigr), G\bigl(\mathbf{R}_j^{\text{UE}_k}\bigr)\bigr)\bigr) + \alpha L_2 \bigr]
			\end{equation}
			}
			
		}
	\textbf{Output} the trained SE-CGAN, $G(\bar{\mathbf{R}}^{\text{BS}}; \hat{\Lambda}_{g})$, at the ISAC BS and the trained CE-CGAN, 	$G(\bar{\mathbf{R}}^{\text{UE}_{k}}; \hat{\Lambda_{g}})$, at the downlink UE$_k$.
\end{algorithm}

%			\[
%			\nabla_{\Lambda_g} \frac{1}{b} \sum_{j=1}^b \log (1 - D(G(z^{(i)}, y^{(i)})))
%			\]

%	\[
%			\nabla_{\Lambda_d} \frac{1}{b} \sum_{j=1}^b [\log D(x^{(i)}, y^{(i)}) + \log (1 - D(G(z^{(i)}, y^{(i)})))]
%			\]

				\def\leftvspace{\rule{0pt}{4.5mm}}
\def\rightvspace{2mm}

\begin{table*}[h!]
	\centering
	\caption{Architecture of SE-CGAN and CE-CGAN.}
	\label{tab:network_architecture}
	\begin{tabular}{c|c|c|c|c|c}
		%		\toprule
		\textbf{Model} &  \textbf{Network} &	\textbf{Layers} & \textbf{Size}  & \textbf{Filter} & \textbf{Activation Function} \\  \hline
		\multirow{8}{*}{SE-CGAN} & \multirow{4}{*}{Generator} &  Input  & $2M P$ & \centering - & - \\
		& & FFL    & $100$    & \centering - & {LeakyReLU} \\
		& & FFL    & $200$     & \centering - & {LeakyReLU} \\
		& & Output & $2M^2$  & \centering - & - \\  \cline{2-6}
		
		& \multirow{4}{*}{Discriminator} &  Input  & $2M^2$ & \centering - & - \\
		& & FFL    & $100$    & \centering - & {LeakyReLU} \\
		& & FFL    & $200$     & \centering - & {LeakyReLU} \\
		& & Output & $1$  & \centering - & - \\ \hline

		\multirow{8}{*}{CE-CGAN} & \multirow{4}{*}{Generator}  & Input  & $2 P$ & \centering - & - \\
		& & CL    & $132$     & \centering $4$ & {LeakyReLU} \\
		& & FFL    & $500$     & \centering - & {LeakyReLU} \\
		& & Output & $2M N$  & \centering- & - \\ \cline{2-6}
		& \multirow{4}{*}{Discriminator}  & Input  &  $2M N$ & \centering - & - \\
		& & CL    & $132$    & \centering $4$ & {LeakyReLU} \\
		& & FFL    & $500$     & \centering - & {LeakyReLU} \\
		& & Output & $1$  & \centering- & - \\ \hline
		
	\end{tabular}
\end{table*}

		\vspace{-2mm}

%__________________________________________________________________
\section{Complexity Analysis}~\label{Sec:Complexity_Analysis}

	{Based on the proposed SE-CGAN and CE-CGAN {channel} estimation frameworks, this section derives  {the computational complexity required by the trained CGAN-based framework to estimate the SAC channel. The computational complexity is computed in terms of} {the} number of real additions, $C_\mathcal{A}$, and multiplications, $C_\mathcal{M}$. To provide a comprehensive analysis, we first consider the operations involved in the generator and discriminator networks of the proposed SE-CGAN. Both networks consist of several FFL, each contributing to the overall computational load. In particular, the proposed SE-CGAN contains two hidden FFL and an input and output layers. Let $\eta_i$ be the number of neurons of the $i$-th layer. {Therefore, according to the parameters presented in Table \ref{tab:network_architecture}, the computational complexity  introduced by the first FFL is $\eta_{2}^{\text{SE}} (\eta_1^{\text{SE}} + 1)$  real additions and $\eta_1^{\text{SE}} \eta_{2}^{\text{SE}}$ real multiplications.} Similarly, the rest of the FFLs have the complexity corresponding to the number of neurons. To this end, the required number of additions and multiplications required by the generator and discriminator networks of SE-CGAN are respectively given as}

	\begin{align}
		& C^{\text{SE}}_{\mathcal{A}_G} = \sum_{i=1}^3 \eta_i^{\text{SE}} \eta_{i+1}^{\text{SE}} + \sum_{i=1}^3 \eta_{i+1}^{\text{SE}}, \\
		& C^{\text{SE}}_{\mathcal{M}_G} = \sum_{i=1}^3 \eta_i^{\text{SE}} \eta_{i+1}^{\text{SE}},
	\end{align}
	
	\noindent and 
	
	\begin{align}
		& C_{\mathcal{A}_D}^{\text{SE}} = \sum_{i=1}^2 \eta_i^{\text{SE}} \eta_{i+1}^{\text{SE}} + \sum_{i=1}^2 \eta_{i+1}^{\text{SE}} + (\eta_{3}^{\text{SE}} + 1), \\
		& C_{\mathcal{M}_D}^{\text{SE}} = \sum_{i=1}^2 \eta_i^{\text{SE}} \eta_{i+1}^{\text{SE}} + \eta_{3}^{\text{SE}}.
	\end{align}

	To this end, the overall necessary number of real additions and multiplications of the proposed SE-CGAN is represented as
	
	\begin{align}
		& C_{\mathcal{A}}^{\text{SE}} =  2\eta_1^{\text{SE}} \eta_2^{\text{SE}} + \sum_{i=2}^3 k_i \eta_i^{\text{SE}} + \eta_{4_G}^{\text{SE}} + 1, \\
		& C_{\mathcal{M}}^{\text{SE}} = \sum_{i=1}^2 2 \eta_i^{\text{SE}} \eta_{i+1}^{\text{SE}} + \eta_{3}^{\text{SE}} (\eta_{4_G}^{\text{SE}} + 1).
	\end{align}
	
	\noindent where $k_2 = 2(\eta_{3}^{\text{SE}} + 1)$ and $k_3 = 3 + \eta_{4_G}^{\text{SE}}$.

	{On the other hand, the CE-CGAN network contains a CL, FFL, input and output layers. Similarly, the computational complexity contributions arise from both generator and discriminator networks. To this end, by summing up the contributions from all layers in both the generator and discriminator networks, the total computational complexity of the proposed CE-CGAN framework can be obtained. Let $F_z$ {be} the filter size, $F_n$ {be} the number of filters, and $F_s$ {be} the stride. The output size of a CL is given by}

	\begin{equation}
		\eta_F = \lfloor{\frac{\eta_1^{\text{CE}} -  F_z}{F_s}+1\rfloor},
	\end{equation}

	\noindent {where $\lfloor{\cdot\rfloor}$  is the floor operation. The computational complexity is introduced by the the second, third, and fourth layers, since the flatten layer does not contain any addition or multiplication complexity. The number of required additions  is presented as $(F_z+1)\eta_F F_n$, $\eta_3^{\text{CE}}(\eta_F F_n+1)$ and $(\eta_4^{\text{CE}}+1)\eta_3^{\text{CE}}$, respectively, while the number of required multiplications is presented as $F_z\eta_F F_n$, $\eta_3^{\text{CE}}\eta_F F_n$, and $\eta_3^{\text{CE}} \eta_4^{\text{CE}}$, respectively.  To this end, the required number of additions and multiplications required by the generator and discriminator networks of CE-CGAN are respectively given as}

	\begin{align}
		& C_{\mathcal{A}_G}^{\text{CE}} = (F_z+\eta_3^{\text{CE}}+1)\eta_F F_n +(\eta_{4_G}^{\text{CE}}+1)\eta_3^{\text{CE}} + \eta_{4_G}^{\text{CE}}, \\
		& C_{\mathcal{M}_G}^{\text{CE}} = (F_z+\eta_3^{\text{CE}})\eta_F F_n +\eta_3^{\text{CE}} \eta_{4_G}^{\text{CE}},
	\end{align}
	
	\noindent and
	
	\begin{align}
		& C_{\mathcal{A}_D}^{\text{CE}} = (F_z+\eta_3^{\text{CE}}+1)\eta_F F_n +2 \eta_3^{\text{CE}} + 1, \\
		& C_{\mathcal{M}_D}^{\text{CE}} = (F_z+\eta_3^{\text{CE}})\eta_F F_n +\eta_3^{\text{CE}}.
	\end{align}

	\noindent Consequently, the total computational complexity of the CE-CGAN framework is {expressed} as 
	
	\begin{align}
		& C_{\mathcal{A}}^{\text{CE}} = 2 (F_z+\eta_3^{\text{CE}}+1)\eta_F F_n + \eta_{4_G}^{\text{CE}} (\eta_3^{\text{CE}} + 1) + 3 \eta_3^{\text{CE}} + 1, \\
		& C_{\mathcal{M}}^{\text{CE}} = 2(F_z+\eta_3^{\text{CE}})\eta_F F_n +\eta_3^{\text{CE}}(\eta_{4_G}^{\text{CE}} + 1).
	\end{align}

	\noindent {To this end, the total number of additions and multiplications required by the proposed CGAN framework is $C_\mathcal{A}^{\text{SE}} + C_\mathcal{A}^{\text{CE}}$ and $C_\mathcal{M}^{\text{SE}} + C_\mathcal{M}^{\text{CE}}$, respectively. In particular, {the number of additions and multiplications is} respectively given as in \eqref{eq:total_A} and \eqref{eq:total_M}.}
	
	\begin{figure*}[!b]
		\begin{equation}
			\label{eq:total_A}
			C_\mathcal{A}^{\text{Total}} = 2\eta_1^{\text{SE}} \eta_2^{\text{SE}} + \sum_{i=2}^3 k_i \eta_i^{\text{SE}} + \eta_{4_G}^{\text{SE}} + 1 + 2 (F_z+\eta_3^{\text{CE}}+1)\eta_F F_n + \eta_{4_G}^{\text{CE}} (\eta_3^{\text{CE}} + 1) + 3 \eta_3^{\text{CE}} + 1,
		\end{equation}
	\end{figure*}
	
	\begin{figure*}[!b]
		\begin{equation}
			\label{eq:total_M}
			C_\mathcal{M}^{\text{Total}} =  \sum_{i=1}^2 2 \eta_i^{\text{SE}} \eta_{i+1}^{\text{SE}} + \eta_{3}^{\text{SE}} (\eta_{4_G}^{\text{SE}} + 1) + 2(F_z+\eta_3^{\text{CE}})\eta_F F_n +\eta_3^{\text{CE}}(\eta_{4_G}^{\text{CE}} + 1.
		\end{equation}
	\end{figure*}

%\pagebreak
	
%	\begin{align}
%		& C_\mathcal{A}^{\text{Proposed}} = C_\mathcal{A}^{\text{SE}} + C_\mathcal{A}^{\text{CE}}, \\
%		& C_\mathcal{M}^{\text{Proposed}} = C_\mathcal{M}^{\text{SE}} + C_\mathcal{M}^{\text{CE}}.
%	\end{align}

			\vspace{-3mm}
	
%__________________________________________________________________
\section{Simulation Results}~\label{Sec:Simulation_Results}

			\vspace{-3mm}
This section extensively validates the performance of the proposed CGAN {channel} estimation framework for the RIS-assisted ISAC system. First, the simulation parameters and setup are presented. Then, the SAC channel estimation performance is evaluated under different SNR {values} and system conditions. 
%------------------------------------------------------------------
	\subsection{Simulation Parameters}
		Let $K = 3$, $M = 4$, and $N = 30$ for all the following simulations, unless further specified. 
%		\textcolor{blue}{Referring to the 3rd Generation Partnership Project standard, the duration of each time slot (i.e., pilot symbol duration) is defined as $T_P = \unit[0.52]{\mu s}$ \cite{3gpp2016eutra}. As such, the sub-frame duration in the proposed pilot transmission policy can be devised as $T_F = T_P P$. With $P = M$, the sub-frame duration in obtained as $T_F = \unit[2.08]{\mu s}$. Correspondingly, the total sub-frame duration required for the estimation is $T_E = C T_F = \unit[62.4]{\mu s}$. According to \cite{coh}, the channel coherence time is set to $T_{\text{coh}} = \unit[1]{ms}$, which indicates that $T_E$ is much smaller than $T_{\text{coh}}$. This implies that by adopting the proposed pilot transmission policy, the training overhead of the proposed estimation approach is acceptable, satisfying the low-cost requirements of both ISAC BS and downlink UEs in real-world systems.} 
		The sensing channel is modeled according to the radar channel model as \cite{radar1, radar2}
		
		\begin{equation}
			\mathbf{A} = {\mu} \mathbf{a}(\theta) \mathbf{a}(\theta)^H.
		\end{equation}
		\noindent Here, {$\mu$} denotes the complex-valued reflection coefficient associated with the target with phase shifts uniformly distributed from $[0, 2 \pi)$, and $\mathbf{a}(\theta)$ is the steering vector, expressed as
		
		\begin{equation}
			\label{Eq: steering}
			\mathbf{a}(\theta) = \left[1, e^{j \frac{2\pi d}{\lambda} \sin(\theta)}, \ldots, e^{j \frac{2\pi d (M-1)}{\lambda} \sin(\theta)}\right]^T,
		\end{equation}

		 \noindent where {$\theta = - \frac{2 \pi}{3}$}, $d$, and $\lambda$ denote the azimuth angle, BS antenna spacing, and signal wavelength, respectively. On the other hand, the communication channels (i.e., $\mathbf{H}$ and $\mathbf{r}_k$) are modeled as Rician, {being} expressed as

		 \begin{equation}
		 	\mathbf{h} = \sqrt{\text{PL}} \left(\frac{K_1}{K_1 + 1}\mathbf{\bar{h}} +  \frac{1}{K_1 + 1}\mathbf{\tilde{h}}\right),
		 \end{equation}

		 \noindent where $\mathbf{h}$ represents the channels (i.e., $\mathbf{h} = \{\mathbf{H}, \mathbf{r}_k\}$), {PL denotes the path loss}, and $K_1$ is the Rician factor. $\mathbf{\bar{h}}$ and $\mathbf{\tilde{h}}$ {are the line-of-sight and non-line-of-sight} components of the channel, respectively. Here, $\mathbf{\bar{h}} = \mathbf{a}(\dot{\theta}) \mathbf{a}(\bar{\theta})^H$, where $\mathbf{a}(\dot{\theta})$ corresponds to the angle of departure from the source to destination (i.e., BS to RIS/RIS to UE$_{k}$) and $\mathbf{a}(\bar{\theta})$ corresponds to the angle of arrival (i.e., RIS/UE$_{k}$) and are set to $\frac{\pi}{3}$. $\mathbf{\bar{h}}$ is formulated similar to \eqref{Eq: steering}, whereas $\mathbf{\tilde{h}}$ is the random component containing independent and identical distributed $\mathcal{CN}(0, 1)$ elements. The Rician factor, $K_{1}$ is set to $10$ and $0$ (i.e., representing a Rayleigh fading model) for $\mathbf{H}$ and $\mathbf{r}_{k}$, respectively \cite{chan_ref}. Furthermore, {the PL} is modeled as $\text{PL} = \text{PL}_{r} (\frac{d_j}{d_{r}})^{-\zeta_j}$. $d_{j}$ is the distance, $\text{PL}_{r}$ is the path loss at a reference distance ${d_r}$, and $\zeta_j$ is the path loss exponent. We set $PL_r = \unit[-30]{dBm}$ and $ {d_r} = \unit[1]{m}$. The path loss exponents of the BS-target-BS, BS-RIS, and RIS-UE$_{k}$ links are  $\zeta_{1} = 3$, $\zeta_{2} = 2.3$,  and $\zeta_{3} = 2$, respectively, while the distances are set as $d_1 = \unit[140]{m}$, $d_2 = \unit[50]{m}$, and $d_3 = \unit[2]{m}$. According to \cite{Yu_Globecom, P_ref}, the transmit power of the ISAC BS and	is set to $P = \unit[-20]{dBm}$.

		 The dataset {size is} $Q \times V = 10^4$ for each SNR value, {with} $Q = 1000$ and $V = 10$. {In this work,} $90\%$ of the {dataset} size is used for training, while the remaining $10\%$ is used for testing. {The SNR values in the training stage {are} $\text{SNR} =  \unit[10\colon5\colon20]{dB}$, whereas the testing stage {uses} $\text{SNR} = \unit[-10\colon2.5\colon30]{dB}$, which includes $17$ values from $\unit[-10]{dB}$ to $\unit[30]{dB}$ with a step increment of $\unit[2.5]{dB}$. The choice of the SNR region ensures that {the} model encounters instances from unfamiliar conditions, not only unseen samples.} This way, the simulation results confirm the generalization of the model and eliminate the need of estimating the SNR prior to estimating the desired channels. Lastly, the scaling factor, $\rho$, is set to $10^4$.

		 To validate the performance of the proposed CGAN {channel} estimation approach, the normalized mean square error {(NMSE)} is considered as the main {performance} {metric being expressed as}
		 
		 \begin{equation}
		 	\text{NMSE} = \mathbb{E}\biggl\{\frac{\bigl\|\mathbf{h}_{\text{Estimated}} - \mathbf{h}_{\text{True}}\bigr\|^2_{F} }{\bigl\|\mathbf{h}_{\text{True}}\bigr\|^2_F}\biggr\}.
		 \end{equation}
		 
		 \noindent Furthermore, the work in \cite{Yu_Globecom} is considered as a benchmark to effectively evaluate the performance of the proposed algorithm  in the following {subsections.}

%		 The distances
%		 of BS-target-BS, BS-IRS, and IRS-Uk are set to dS = 140m,
%		 dBI = 50m, and dIUk = 2m, respectively. The corresponding
%		 path loss exponents are γS = 3, γBI = 2.3, and γIUk = 2,
%		 respectively [10]. 

%------------------------------------------------------------------
	\subsection{Impact of Varying {the} SNR}
\begin{figure}[!t]
		\centering
		\includegraphics[width=0.51\textwidth]{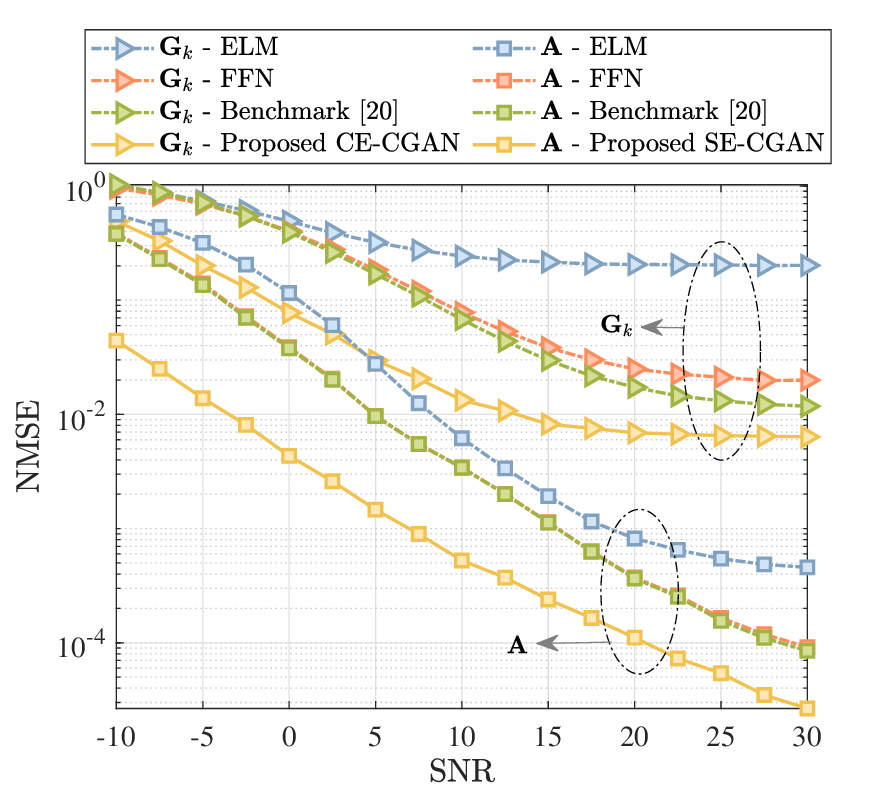}
		\caption{NMSE performance for {estimating} SAC channels.}
		\label{Fig:V_SNR}
	\end{figure}

	{Fig. \ref{Fig:V_SNR}} shows the estimation performance under different SNR conditions. {{The NMSEs} of the proposed CE-CGAN and SE-CGAN demonstrate a considerable improvement compared to the benchmark estimation approach presented in \cite{Yu_Globecom} as well as other traditional models, including a FF network (FFN) and extreme learning machine (ELM).}  {It is worth noting that the FFN model consists of two hidden layers, each containing $256$ neurons, with hyperparameters aligned with the benchmark in \cite{Yu_Globecom}. The ELM model, on the other hand, consists of one hidden layer of $256$ neurons to estimate both channels. These specifications ensure a fair comparison across all approaches.} For the {proposed} SE-CGAN, this improvement is especially notable at lower SNR levels, where the proposed approach {demonstrates} good noise resilience, {thus achieving lower NMSE values}. This is one of the CGAN advantages, where the generator learns to create samples, not to map input data to an output. It generates data conditioned by the input sample but based on the loss function design, {thereby reducing} the noise impact. Furthermore, as the communication {channel} estimation is considered more complicated due to the cascaded relationship in \eqref{Eq:RX_signal_UE_F}, one can observe that the achieved NMSE is {higher} than that of the sensing link. {However, overall, the proposed scheme outperforms the benchmark scheme and the additional traditional models substantially.  {In particular, the proposed approach achieves around $\unit[8]{dB}$ SNR improvement at $\text{NMSE} = 10^{-2}$ and $\text{NMSE} = 10^{-1}$ as compared to \cite{Yu_Globecom} for estimating $\mathbf{A}$ and $\mathbf{G}_{k}$, respectively.}}
%	In particular, at $\text{NMSE} = 10^{-2}$, the proposed approach achieves around $\unit[8]{dB}$ SNR improvement for estimating $\mathbf{A}$ compared to the benchmark scheme. 
%	
%	Moreover, the proposed approach achieves around $\unit[8]{dB}$ SNR improvement at $\text{NMSE} = 10^{-1}$ for estimating $\mathbf{G}_{k}$ compared to the benchmark scheme.
%	
	The figure also highlights the generalization performance of the proposed CGAN approach, where it outperforms the benchmark schemes even at SNR ranges that were excluded in the training process (i.e., $\text{SNR} = \unit[-10\colon5]{dB}$).

%------------------------------------------------------------------
	\subsection{Impact of Varying $M$}

	\begin{figure}[!t]
		\centering
		\includegraphics[width=0.51\textwidth]{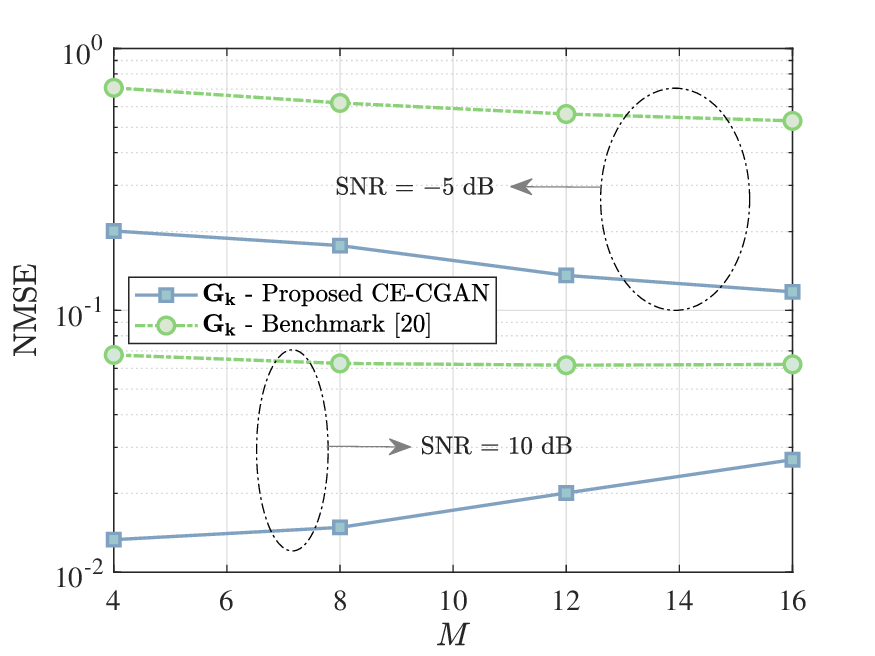}
		\caption{NMSE performance for the communication channel estimation versus $M$ {at $N = 30$}.}
		\label{Fig:V_M_R}
	\end{figure}
	
	 	{Fig. \ref{Fig:V_M_R}}  analyzes the performance of {the} proposed approach in estimating the communication channel, $\mathbf{G}_k$, with respect to  varying $M$ under different SNR conditions: low (i.e., $\unit[-5]{dB}$) and high (i.e., $\unit[10]{dB}$) SNR values. As illustrated, the proposed CGAN consistently outperforms the benchmark scheme across all $M$ values and SNR conditions, demonstrating superior channel estimation accuracy. The CGAN robustness in handling higher dimensions can be attributed to its ability to better model complex channel conditions and effectively learn from the discriminator feedback, even for larger sets of training data. {Furthermore, the decrease in {the} NMSE as $M$ increases at $\text{SNR} = \unit[-5]{dB}$ indicates that the proposed framework is capable of leveraging the additional information {present in the received signals  to refine its estimation}.} {This robustness stems from the ability of CGANs to capture and model the inherent spatial correlations in the communication channel. By leveraging its generation mechanism, the CGAN framework enables more effective utilization of the additional spatial diversity introduced as $M$ increases. This results in enhanced noise suppression at low SNR values.} 
%	 	This is due to the CGAN approach in estimating channels, where it is not just learning static mappings, but it reproduces complex channel characteristics that are conditioned by the given input. 
	 	This is a particularly useful feature for ISAC systems (i.e., where lower SNR values are adopted). At $\text{SNR} = \unit[10]{dB}$, where the noise level is relatively low, {the primary challenge of the proposed CGAN shifts from combating noise to accurately capturing the high-dimensional channel details. The generator network may face difficulties in fully capturing the high-dimensional dependencies, which results in a gradual increase in estimation error, especially that the overall network architecture and hyperparameters are optimized at lower channel dimensions (i.e., $M=4$ and $N = 30$).}  {This slight degradation at higher SNR can be attributed to increased sensitivity to minor inaccuracies in capturing subtle channel characteristics. Despite this, the NMSE increase remains minimal, demonstrating the CGAN's effectiveness in maintaining reliable performance. Moreover, the proposed approach consistently outperforms {the previous work}, showcasing its adaptability and robustness across diverse real-world conditions.}

%\textcolor{blue}{This slight degradation at higher SNR can be attributed to increased sensitivity to minor inaccuracies in capturing subtle channel characteristics. Despite this, the NMSE increase remains minimal, demonstrating the CGAN's effectiveness in maintaining reliable performance. Moreover, the proposed approach consistently outperforms {the previous work}, showcasing its adaptability and robustness across diverse real-world conditions.}

	 \begin{figure}[!t]
	 	\centering
	 	\includegraphics[width=0.51\textwidth]{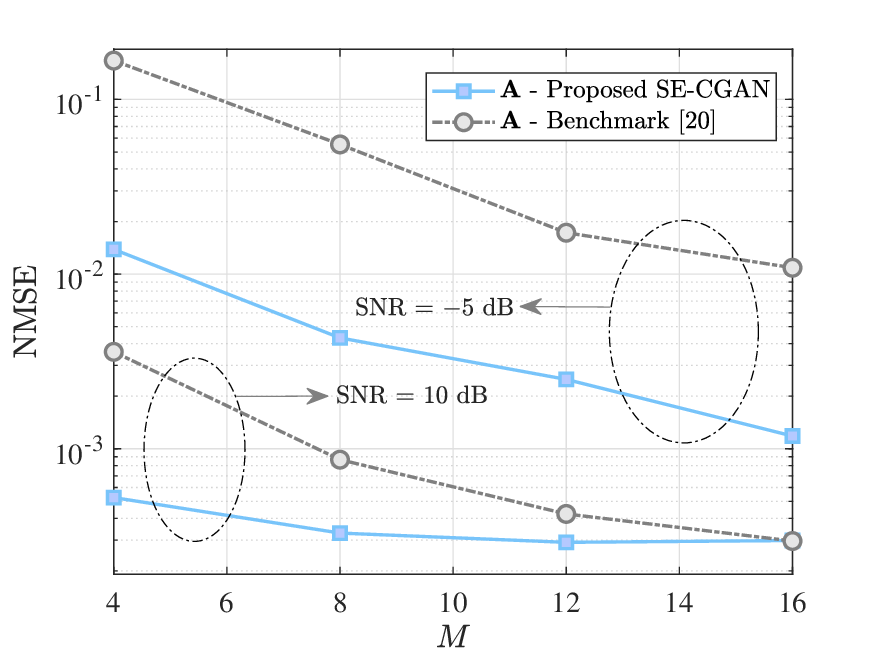}
	 	\caption{NMSE performance for the sensing channel estimation versus $M$ {at $N = 30$}. \vspace{-4mm}}
	 	\label{Fig:V_M_D}
	 \end{figure}

		{Fig. \ref{Fig:V_M_D}}  demonstrates the performance of {the} proposed approach in estimating the sensing channel, $\mathbf{A}$, with respect to  varying $M$ {under the} same SNR conditions considered in Fig. \ref{Fig:V_M_R}. It can be seen that the proposed SE-CGAN algorithm significantly outperforms the benchmark {in \cite{Yu_Globecom}} at both SNR levels, evidencing its enhanced ability to estimate sensing channels under different channel dimension and noise conditions. {At the SNR of $\unit[-5]{dB}$}, the NMSE of both models decreases, yet the {proposed} SE-CGAN maintains a stable  performance enhancements, showcasing its robustness against high noise levels. The consistent performance across increasing $M$ values suggests that the {proposed} SE-CGAN effectively utilizes additional antennas to mitigate noise through its adaptive learning and channel modeling capabilities. On the other hand, {at the SNR of $\unit[10]{dB}$}, although the NMSE for the benchmark model decreases with higher $M$, the proposed algorithm still holds a significant advantage. In particular, it begins with an outperforming estimation for smaller number of antennas, demonstrating robust performance right from the {onset}. The performance of the proposed approach saturates as $M$ increases, which proves its scalability and capability to handle high-dimensional channel environments without substantial loss in performance. At $M = 16$, both models converge to similar levels of NMSE, illustrating that while the benchmark {in \cite{Yu_Globecom}} improves, the {proposed} SE-CGAN effectively sustains its superior channel estimation capabilities across varying antenna configurations. This balance of initial superiority and scalability at high SNR settings emphasizes the {proposed SE-CGAN} advantage in complex RIS-assisted ISAC systems where both accuracy and adaptability are crucial.

			\vspace{-3mm}
	
%------------------------------------------------------------------
	\subsection{Impact of Varying $N$}

	\begin{figure}[!t]
		\centering
		\includegraphics[width=0.51\textwidth]{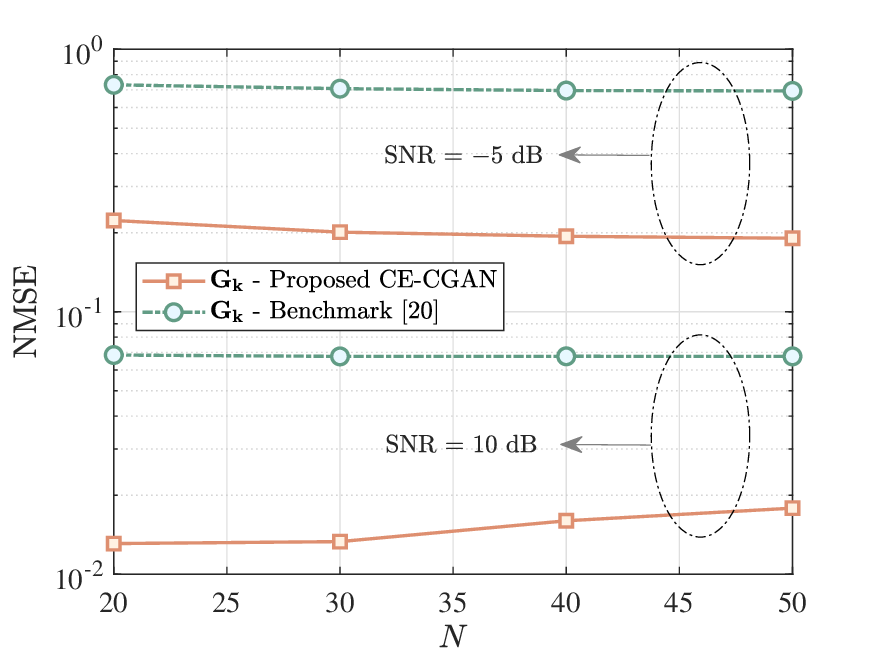}
		\caption{NMSE performance for the communication channel estimation versus $N$ {at $M = 4$}. \vspace{-4mm}}
		\label{Fig:V_N_R}
	\end{figure}

			{Fig. \ref{Fig:V_N_R}}  demonstrates the performance of {the} proposed approach in estimating the communication channel, $\mathbf{G}_k$, with respect to  varying $N$ under the same SNR conditions considered throughout the simulations. Similarly, the proposed CGAN algorithm effectively  outperforms the benchmark scheme {in \cite{Yu_Globecom}} across all $N$ values and SNR conditions. At lower SNR (i.e., $\text{SNR} = \unit[-5]{dB}$), the proposed algorithm demonstrates {a substantial} improvement over the benchmark, with the NMSE slightly decreasing as $N$ increases. This significant reduction emphasizes the CGAN robustness in noisy environments, effectively leveraging the increased number of RIS elements to enhance channel estimation. Alternatively, at higher SNR (i.e., $\text{SNR} = \unit[10]{dB}$), although the CGAN continues to outperform the benchmark, the performance improvement decreases, indicating less gains from additional RIS elements. {However, it is worth noting that the NMSE remains relatively stable even with increasing $N$, reflecting the CGAN ability to maintain its estimation accuracy across larger RIS setups. The slight increase in NMSE at higher SNR is marginal and does not significantly impact the estimation performance.}  Overall, the performance proves the CGAN scalability to larger configurations and shows the CGAN proficiency in utilizing the spatial diversity offered by larger RIS setups to optimize channel estimation, making it particularly useful for ISAC systems operating across a range of SNR scenarios.

			{Fig. \ref{Fig:V_N_R}} demonstrates the performance of {the} proposed approach in estimating the communication channel, $\mathbf{G}_k$, with respect to varying $N$ under the same SNR conditions considered throughout the simulations. Similarly, the proposed CGAN algorithm effectively outperforms the benchmark scheme {in \cite{Yu_Globecom}} across all $N$ values and SNR conditions. At lower SNR (i.e., $\text{SNR} = \unit[-5]{dB}$), the proposed algorithm demonstrates {a substantial} improvement over the benchmark, with the NMSE slightly decreasing as $N$ increases. This significant reduction emphasizes the CGAN robustness in noisy environments, effectively leveraging the increased number of RIS elements to enhance channel estimation. Alternatively, at higher SNR (i.e., $\text{SNR} = \unit[10]{dB}$), although the CGAN continues to outperform the benchmark, the performance improvement decreases, indicating less gains from additional RIS elements. {However, it is worth noting that the NMSE remains relatively stable even with increasing $N$, reflecting the CGAN ability to maintain its estimation accuracy across larger RIS setups. The slight increase in NMSE at higher SNR is marginal and does not significantly impact the estimation performance.} {It is also important to note that the model was optimized for a baseline configuration (i.e., $M=4$ and $N = 30$), which explains the observed behavior when scaling beyond these dimensions.} Overall, the performance proves the CGAN scalability to larger configurations and shows the CGAN proficiency in utilizing the spatial diversity offered by larger RIS setups to optimize channel estimation, making it particularly useful for ISAC systems operating across a range of SNR scenarios. 
		\vspace{-3mm}
%		, demonstrating the model adaptability and resilience in handling high-dimensional channels without considerable performance degradation
		
%------------------------------------------------------------------
%	\subsection{Convergence}

%------------------------------------------------------------------
	\subsection{Complexity Evaluation}
		\begin{figure}[!t]
		\centering
		\subfloat[]{
			\includegraphics[width=0.45\textwidth]{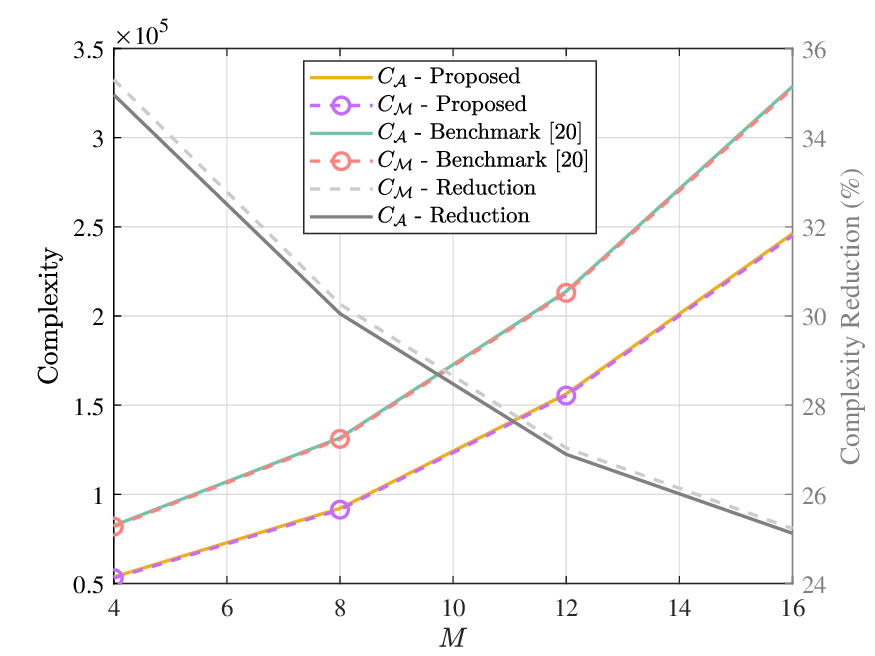}
			\label{Fig:Comp_D}
			\vspace{-4mm}
		}
		\hfil
		\subfloat[]{
			\includegraphics[width=0.45\textwidth]{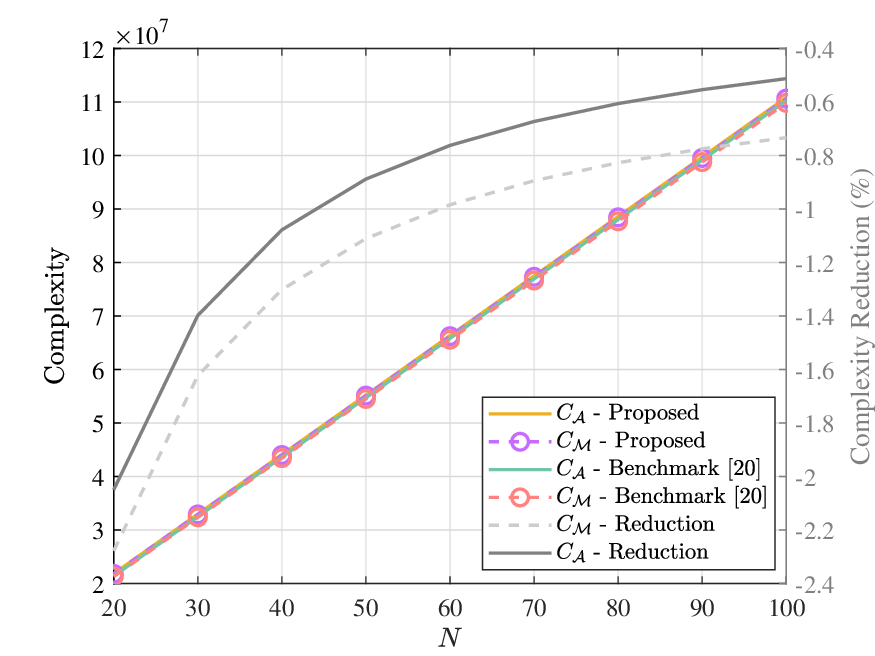}
			\label{Fig:Comp_R}
			\vspace{-4mm}
		}
		\caption{Computational complexity evaluation. (a) Sensing link, (b) Communication link. \vspace{-4mm}}
		\label{Fig:Comp}
	\end{figure}

		{Figs. \ref{Fig:Comp_D} and \ref{Fig:Comp_R}} {show} the computational complexity of the proposed CGAN approach as compared to the benchmark scheme presented in \cite{Yu_Globecom}. The complexity is analyzed in terms of the required number of real additions and multiplications according to the derived formulations in Section \ref{Sec:Complexity_Analysis}. Furthermore, the figures show the complexity reduction of using the proposed CGAN algorithm over the benchmark scheme, which is expressed as

	\begin{equation}
		\text{Reduction} =  \frac{ C^{\text{\cite{Yu_Globecom}}}_\chi  - C^{\xi}_\chi }{C^{\text{\cite{Yu_Globecom}}}_\chi  }, \hspace{0.2em} \chi \in \{\mathcal{A}, \mathcal{M}\}, \xi \in \{\text{SE}, \text{CE}\}.
	\end{equation}
	
	\noindent 	{Fig. \ref{Fig:Comp_D}} illustrates the computational complexity and reduction percentage of estimating the direct channel (i.e., $\mathbf{A}$) as $M$ increases. As can be seen, the complexities of additions and multiplications for both the proposed and benchmark schemes show a trend of increasing with $M$. However, the proposed algorithm complexity is notably lower compared to the benchmark method. The reduction in complexity, depicted on the secondary {(i.e., right)} y-axis, emphasizes the efficiency gains from the proposed approach as $M$ increases. The proposed {SE-CGAN approach} not only significantly reduces {the} computational complexity compared to the benchmark, enhancing system efficiency, but also delivers a superior performance across various conditions, making it an exceptionally effective solution for ISAC RIS-assisted systems.

	 {Fig. \ref{Fig:Comp_R}} illustrates the computational complexity and reduction percentage of estimating the communication channel (i.e., $\mathbf{G}_k$) as $N$ increases. Similarly, the number of additions and multiplications for both the proposed and benchmark schemes rapidly increase as {$N$} increases.  The computational complexity of the proposed CE-CGAN approach is comparable to the benchmark scheme, even when the channel estimation dimension enlarges. This is due to the fact that estimating the communication channel is challenging and often require complex models.  Despite the inherent challenges, our approach achieves a similar level of computational complexity to the benchmark, while significantly outperforming it in terms of performance. {This proves  the effectiveness of our model in handling the delicate features of cascaded communication channels estimation within RIS-assisted ISAC systems, ensuring practicality and timeliness in real-world deployments.}

%	 		\textcolor{blue}{Fig. 7b illustrates the computational complexity and reduction percentage of estimating the communication channel (i.e., $\mathbf{G}_k$) as $N$ increases. Similarly, the number of additions and multiplications for both the proposed and benchmark schemes rapidly increase as {$N$} increases.  The computational complexity of the proposed CE-CGAN approach is comparable to the benchmark scheme, even when the channel estimation dimension enlarges. This is due to the fact that estimating the communication channel is challenging and often require complex models.  Despite the inherent challenges, our approach achieves a similar level of computational complexity to the benchmark, while significantly outperforming it in terms of performance. This proves  the effectiveness of our model in handling the delicate features of cascaded communication channels estimation within RIS-assisted ISAC systems, ensuring practicality and timeliness in real-world deployments.}

		\vspace{-3mm}
%__________________________________________________________________
\section{Conclusion} ~\label{Sec:Conclusion}
	This paper {has} investigated the channel estimation problem of an RIS-assisted ISAC system. A novel CGAN approach {has been} proposed to enhance the estimation accuracy and stability. The proposed method {has leveraged} the adversarial training of two deep learning networks to accurately estimate {the} channel {coefficients}, demonstrating a superior performance over conventional techniques. The numerical results {have validated} the efficiency of the proposed approach across different SNR conditions and system dimensions, highlighting its robustness and adaptability. Furthermore, the complexity of the proposed approach {has been} analyzed and compared to {that of the benchmark scheme}. In particular, the {proposed} CE-CGAN model maintains a computational complexity {comparable} to {that of} existing DL methods while achieving {a better} estimation accuracy, while the SE-CGAN model outperforms the existing estimation {model} in both performance and {computational} complexity. This makes the proposed CGAN a promising solution for enhancing the reliability and efficiency of RIS-assisted ISAC systems. {Future works could explore extensions of the proposed CGAN-based channel estimation approach to address further practical challenges. Examples include accommodating multi-target scenarios along with mobility of users and targets, and adapting the framework for wideband system deployments.}

%	These include handling imperfect synchronization between the SAC and residual SI signals, integrating non-linear models for residual SI, accommodating multi-target scenarios, mobility of users and targets, and adapting the framework for wideband system deployments.

	\vspace{-3mm}
%__________________________________________________________________
%\pagebreak
\bibliographystyle{IEEEtran}
{\bibliography{IEEEabrv, References}}

\end{document}